\newcommand{\bld}[1]{\mbox{\boldmath$#1$\unboldmath}}
\newcommand\lmhd{{{\rm L}_{\rm MHD}}}
\newcommand\lpar{{\Lambda_\parallel}}
\newcommand\lout{{\rm L}_{\rm out}}
\newcommand\lcool{{\rm L}_{\rm cool}}
\newcommand\lmfp{{\rm L}_{\rm mfp}}
\newcommand\led{{\rm L}_{\rm ed}}
\newcommand\lequ{{\rm L}_{\rm eq}}
\newcommand\lpd{{\rm L}_{\rm pd}}
\newcommand\lpdperp{{\rm L}_{\rm pd}^{(\perp)}}

\newcommand\cm{\,{\rm cm}}

\def\refnew#1{(\ref{#1})}

\documentclass[preprint]{aastex}
\usepackage{emulateapj5}

\begin{document}
\title{Compressible MHD Turbulence in Interstellar Plasmas}
\author{Yoram Lithwick and Peter Goldreich}
\affil{130-33 Caltech, Pasadena, CA 91125; yoram@tapir.caltech.edu,
pmg@gps.caltech.edu}

\begin{abstract}
Radio-wave scintillation observations reveal a nearly Kolmogorov spectrum
of density fluctuations in the ionized interstellar medium. 
Although this density spectrum is 
suggestive of turbulence, no theory relevant 
to its interpretation exists.
We calculate the density spectrum in 
turbulent magnetized plasmas
by extending the 
theory of incompressible magnetohydrodynamic (MHD) turbulence given by 
\citet{gs95} to include the 
effects of compressibility and particle transport. Our most important results 
are as follows.

\noindent (1) Density fluctuations are due to the slow mode and
the entropy mode. Both modes are passively mixed by the cascade of 
shear Alfv\'en waves. Since the shear Alfv\'en waves have a Kolmogorov spectrum,
so do the density fluctuations.

\noindent (2) Observed density fluctuation amplitudes
constrain the nature of MHD 
turbulence in the interstellar medium. Slow mode density fluctuations are 
suppressed when 
the magnetic pressure is less than the
gas pressure. Entropy mode density fluctuations are suppressed 
by cooling when the cascade timescale
is longer than the cooling timescale. 
These constraints imply 
either that the magnetic and gas pressures are comparable, or that the 
outer scale of the turbulence is very small.

\noindent (3) A high degree of ionization is 
required for the cascade to survive 
damping by neutrals and thereby to extend to small lengthscales. Regions that are
insufficiently ionized produce density fluctuations only on lengthscales larger 
than the neutral damping scale. These regions may account for the excess of power that is
found on large scales.

\noindent (4) Provided that the thermal pressure exceeds 
the magnetic pressure, both the entropy mode and 
the slow mode are damped on lengthscales
below that at which protons can diffuse across an eddy during the eddy's 
turnover time. Consequently, eddies whose 
extents {\it along the magnetic field} 
are smaller than the proton collisional 
mean free path do not contribute 
to the density spectrum. However, in MHD turbulence eddies are highly elongated 
along the magnetic field. From an observational perspective, the relevant 
lengthscale is that {\it transverse} to the magnetic field. Thus the cut-off 
lengthscale for density fluctuations is significantly smaller than the proton 
mean free path.  

\noindent (5) The Alfv\'en mode is critically damped at the transverse 
lengthscale of the proton gyroradius, and thus cascades to smaller lengthscales 
than either the slow mode or the entropy mode.
\end{abstract}
\section {INTRODUCTION}

Diffractive scintillations of small angular-diameter radio sources indicate
that the interstellar electron density spectrum on lengthscales 
$10^8-10^{10}\cm$ is nearly Kolmogorov; i.e. r.m.s. density fluctuations across 
a lengthscale $\lambda$ are nearly proportional to $\lambda^{1/3}$. They also
establish that there are large variations in the amplitude of the density 
spectrum along different lines of sight.

Rickett (1977, 1990) and \citet{ars95}
review the observations of diffractive scintillation and their
interpretation. 
They
also discuss refractive scintillations and dispersion measure fluctuations,
which probe density fluctuations on scales larger than the diffractive
scales.  Non-diffractive measurements tend to indicate that
the Kolmogorov spectrum extends to much larger scales.  However, we focus
primarily on diffractive measurements because they are much more 
sensitive.

Density fluctuations that obey the Kolmogorov scaling occur in homogeneous 
subsonic {\em hydrodynamic} turbulence. They are due to the entropy mode, a 
zero-frequency isobaric mode whose density fluctuations are offset by 
temperature fluctuations. Since subsonic turbulence is nearly incompressible, 
the velocity fluctuations follow Kolmogorov's scaling. To a good approximation, 
the entropy mode is passively mixed by the velocity field, so it also conforms
to the Kolmogorov spectrum.\footnote{Density fluctuations due to the Reynolds 
stress scale as $\lambda^{2/3}$.} Density fluctuations 
in the Earth's atmosphere, 
which cause stars to twinkle, obey the Kolmogorov
scaling. They arise from the passive mixing of the entropy mode. 

The electron density spectrum in the interstellar medium cannot be explained by 
hydrodynamic turbulence.\footnote{Charge neutrality is maintained on 
diffractive 
scales, so electron density fluctuations include compensating fluctuations in 
the density of positive ions.} Because the medium is ionized, magnetic effects
must be accounted for. This is evident since the lengthscales probed by
diffractive scintillations are smaller than the collisional mean free paths of 
both electrons and protons. 
If the magnetic field were negligible,
freely streaming plasma would wipe out density fluctuations 
at diffractive scales.
In the presence of a magnetic field, electrons and protons 
are tied to fieldlines at
the scale of their gyroradii.
For typical interstellar field strengths, these gyroradii are smaller than the 
diffractive scales. A magnetic field thus impedes the plasma from streaming
across fieldlines,
and allows the turbulent cascade and the associated density 
fluctuations to reach very small scales across the fieldlines 
before dissipating. 
Therefore, a 
theory for compressible turbulence in magnetized plasmas is required
to explain the observed density spectra.
Our objective is to develop this theory.

Until now, the only description of density fluctuations in interstellar plasmas 
was by Higdon (1984, 1986). These papers, while prescient, preceded a 
theory for {\em incompressible} MHD turbulence, and therefore did not 
account for the full dynamics of the cascade. We 
compare Higdon's theory with ours in \S \ref{sec:higdon}.

Our compressible theory extends the theory of incompressible MHD turbulence 
given by \citet{gs95} by including a slightly compressible slow mode and a 
passive entropy mode. We also consider kinetic effects: on sufficiently short 
lengthscales, the mean free paths of the particles are significant, and the 
equations of compressible MHD must be modified. This is 
especially important for damping.

In a future paper II, we apply the theory developed here to estimate
amplitudes for density fluctuations produced in supernova shocks, HII regions, 
stellar winds, and thermally unstable regions. These are then compared to 
scattering measures observed along different lines of sight.

Before considering compressible turbulence, we discuss
incompressible MHD turbulence, focusing on issues that are important
for the compressible case.  

\section {INCOMPRESSIBLE MHD TURBULENCE} \label{sec:incomp}

Goldreich \& Sridhar (1995, 1997)
propose a picture of the dynamics of incompressible strong MHD turbulence
and describe the power spectra of Alfv\'en
waves, slow waves, and passive scalars. We extend their picture to cover additional 
features such as the parallel cascades of both slow waves and passive scalars.
Throughout this paper, ``parallel'' and ``transverse'' refer to the orientation 
relative to the ``local mean magnetic field'', which is the magnetic
field averaged over the scale of interest.
Our discussion of incompressible MHD turbulence, while 
somewhat lengthy, is important for understanding the extension to compressible turbulence 
that follows.

Consider a uniform unperturbed plasma with an embedded magnetic field. 
Turbulence is excited at the MHD outer scale, $\lmhd$, by random 
and statistically isotropic forcing, with
r.m.s. velocity fluctuations 
and r.m.s. magnetic field fluctuations (in velocity units) 
which are comparable 
to the Alfv\'en speed, $v_A$.\footnote{The forcing fluctuations may also be
less $v_A$, in which case $\lmhd$ would be defined as the lengthscale at which 
the fluctuations extrapolate to $v_A$.}

As the turbulence cascades from the MHD outer scale to smaller scales, power 
concentrates in modes with increasingly transverse wave vectors. 
The inertial range velocity spectrum applies to lengthscales 
below $\lmhd$ but above the dissipation scale.  It is anisotropic
and is characterized by
\begin{eqnarray}
v_{\lambda_\perp}&=&v_A\Big({{\lambda_\perp}\over
{\lmhd}}\Big)^{1/3},
\label{eq:spectrum1} \\
\lpar&=&
\lambda_\perp^{2/3}\lmhd^{1/3} \label{eq:spectrum2} \ ;
\end{eqnarray}
the inertial range magnetic field spectrum is identical.
Here $\lambda_\perp$ is the lengthscale transverse to the local mean magnetic
field,
$v_{\lambda_\perp}$ is the r.m.s. velocity fluctuation across $\lambda_\perp$,
and $\lpar$ is the lengthscale parallel to the local mean magnetic field
across which the velocity fluctuation is $v_{\lambda_\perp}$.   
We interpret $\lpar$ as
the elongation along the magnetic field of an ``eddy'' which has a
size $\lambda_\perp$ transverse to the magnetic field; it is
not an independent variable, but is a function of $\lambda_\perp$.
Deep within the inertial range, where $\lambda_\perp\ll\lmhd$,
eddies are highly elongated along the
magnetic field: $\lpar\gg\lambda_\perp$. 
In the following subsections, we explain the physics 
underlying the spectrum, and consider some of the implications.

\subsection{Alfv\'en Wave Spectrum}

Arbitrary disturbances can be decomposed into Alfv\'en waves and slow waves. 
The Appendix summarizes the
properties of these waves in the more general case of compressible MHD.
In incompressible MHD, Alfv\'en waves and slow waves are usually referred to as 
shear-Alfv\'en waves and pseudo-Alfv\'en waves, but the former
designation is more convenient for making the connection 
with compressible MHD.  

Our understanding of the MHD turbulence is based on two facts: 
(i) MHD wave-packets 
propagate at the Alfv\'en speed either parallel or antiparallel to the local 
mean magnetic field; and (ii)
nonlinear interactions are restricted to collisions between 
oppositely directed wave-packets. These facts imply that
in encounters between oppositely directed wave-packets, each wave-packet is
distorted as it follows field lines perturbed by its collision partner. 
A wave-packet cascades when the 
fieldlines that it is propagating 
along have spread by a distance comparable to its transverse size. 

Alfv\'en waves have quasi two-dimensional velocity and magnetic field 
fluctuations which are confined to planes perpendicular to the local mean 
magnetic field. As their more complete name shear-Alfv\'en implies, they 
dominate the shear of the mapping of planes transverse to the local mean 
magnetic field produced by field line wander. Thus, Alfv\'en waves control the 
dynamics of MHD cascades; slow waves may be ignored when considering
the dynamics of Alfv\'en waves.

In strong MHD turbulence the cascade time of an Alfv\'en wave-packet is comparable
to its travel time across the parallel length of a single oppositely directed 
Alfv\'en wave-packet of similar size. \citet{gs95} refer to this balance of timescales
as ``critical balance''. It relates the parallel size of a wave-packet, $\lpar$,
to its transverse size, $\lambda_\perp$. Wave-packets of transverse size 
$\lambda_\perp$ cascade when the fieldlines they follow wander relative to each 
other by a transverse distance $\lambda_\perp$. Critical balance implies that this
occurs over a parallel distance $\lpar$. 

The Alfv\'en wave spectrum is given by equations (\ref{eq:spectrum1}) and 
(\ref{eq:spectrum2}), with  $v_{\lambda_\perp}$ referring to the velocity
fluctuations of the Alfv\'en waves.  It is deduced from two scaling arguments:
(i) Kolmogorov's argument that
the cascade time $t_{\lambda_\perp}\simeq
\lambda_\perp/v_{\lambda_\perp}$ leads
to an energy cascade rate,
$v_{\lambda_\perp}^2/t_{\lambda_\perp}\simeq
v_{\lambda_\perp}^3/\lambda_\perp$, which is independent of
lengthscale; and (ii) the ``critical balance'' assertion that
the linear wave period which characterizes the Alfv\'en waves
in a wave-packet is comparable to the 
nonlinear cascade time of that wave-packet,
i.e. $v_A/\lpar\simeq 1/t_{\lambda_\perp}$.

Before considering slow waves in MHD turbulence, we discuss 
two topics that are governed by the dynamics of Alfv\'en waves only: eddies
and passive scalars.

\subsection{Eddies}
\label{subsec:eddies}

\begin{figure}
\plotone{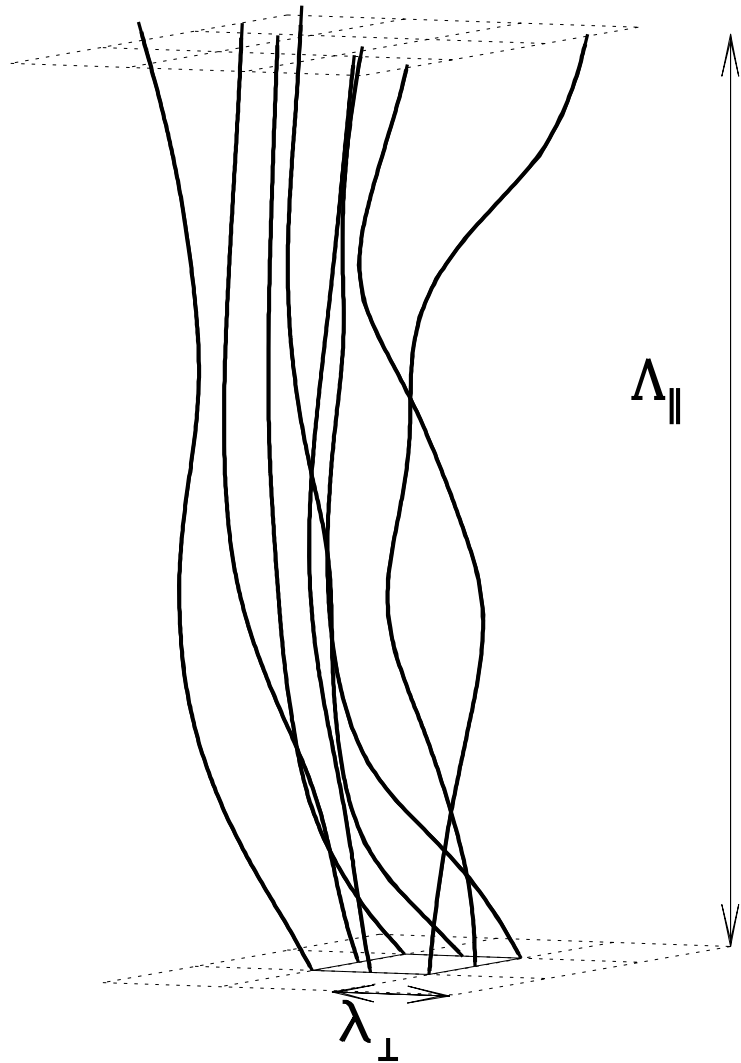}
\caption{\it Wandering of Magnetic Fieldlines:
a fieldline bundle of transverse size $\lambda_\perp$
diverges after a parallel distance $\lpar$, where $\lpar$
is the parallel size of an eddy (eq. [\ref{eq:spectrum2}])
as determined by critical balance.
\label{fig:fieldlines}}
\end{figure}

Because of their transverse polarization,
Alfv\'en waves 
are responsible for the wandering of magnetic fieldlines.
A snapshot of wandering fieldlines is shown in Figure 
\ref{fig:fieldlines}. Each of these fieldlines passes through a localized
region of size $\lambda_\perp$ in one plane transverse to the mean magnetic 
field.  Away from this plane the bundle of fieldlines diverges due to the 
differential wandering of the individual lines. At a second plane,
the bundle's cross sectional area has 
approximately doubled.
Critical balance implies that the distance to this second plane 
is comparable to the parallel wavelength which characterizes the 
bundle, $\lpar$.   
As the bundle spreads, 
other fieldlines, not depicted, enter from its sides. In general, the 
neighboring fieldlines of any individual field line within a region of 
transverse size $\lambda_\perp$ change substantially 
over a parallel distance of 
order $\lpar$. It is natural to think of $\lpar$ as the parallel size of
an ``eddy'' that has transverse size $\lambda_\perp$. 
Two eddies with the same transverse lengthscale that are separated by a 
parallel distance greater than their 
$\lpar$ incorporate different fieldlines, and 
hence are statistically independent. Eddies are distinct from wave-packets. The 
former are rooted in the fluid whereas the latter 
propagate up and down magnetic 
fieldlines at the Alfv\'en speed.

\begin{figure}
\plotone{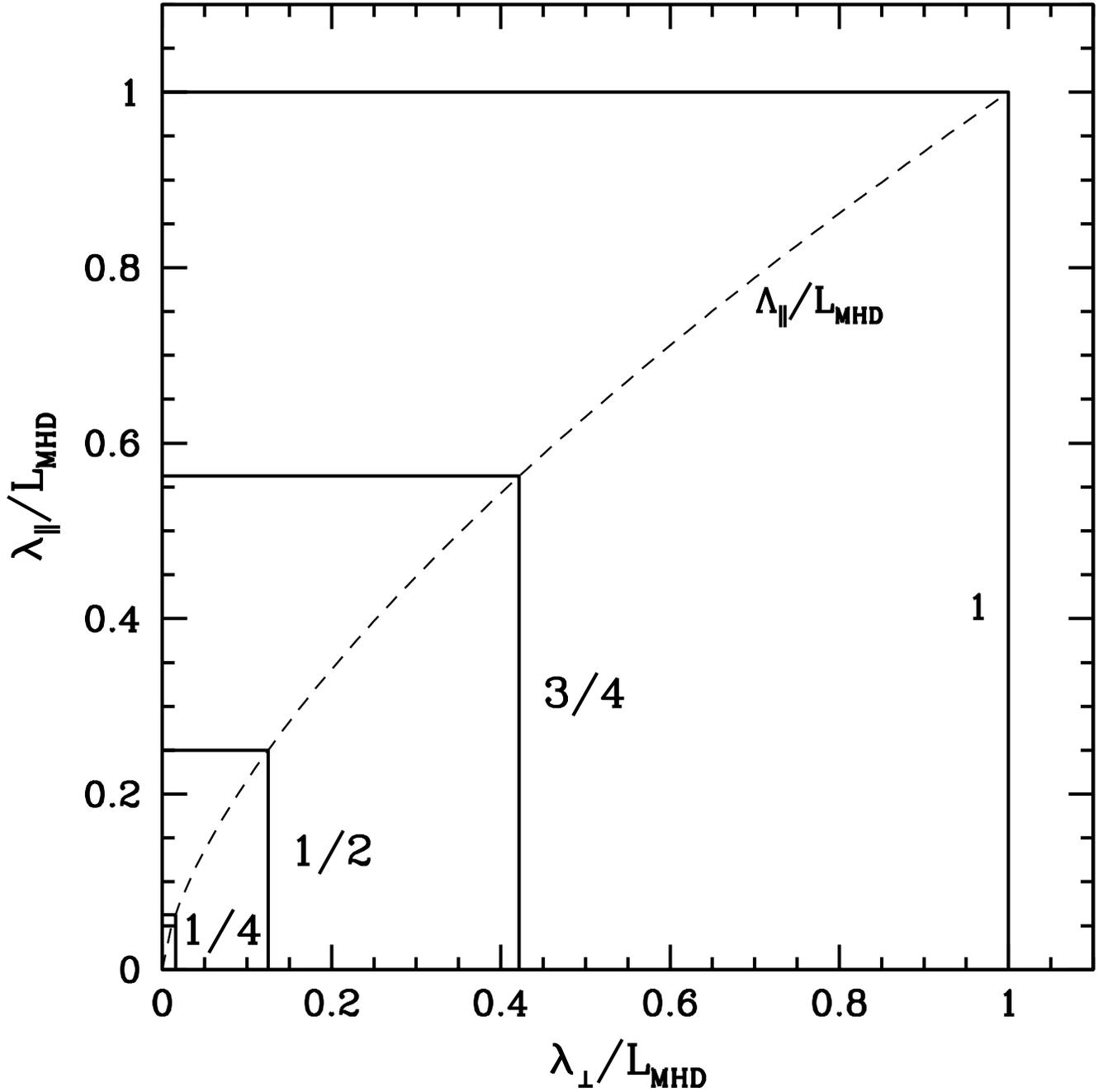}
\caption{\it Three-Dimensional Spectrum: contours of constant
$v_{\lambda_\perp,\lambda_\parallel}$ (eq. [\ref{eq:3dspectrum}]),
labelled by the value of $v_{\lambda_\perp,\lambda_\parallel}/v_A$.
Along the $\lambda_\perp$-axis, $v_{\lambda_\perp,0}/v_A=
(\lambda_\perp/\lmhd)^{1/3}$; along the $\lambda_\parallel$-axis,
$v_{0,\lambda_\parallel}/v_A= (\lambda_\parallel/\lmhd)^{1/2}$.
\label{fig:contours}}
\end{figure}

Aside from their anisotropy, eddies in MHD turbulence are similar to those in 
hydrodynamic turbulence. They are spatially localized structures with 
characteristic velocity fluctuations and lifetimes. The r.m.s. velocity 
difference between two points is determined by the smallest eddy that contains 
both. Different eddies of a given size are statistically independent. The 
three-dimensional spectrum for r.m.s. velocity fluctuations across transverse
lengthscales $\lambda_\perp$ and parallel lengthscales $\lambda_\parallel$ is
\begin{equation}
v_{\lambda_\perp,\lambda_\parallel}=
v_A
\times\left\{
\begin{array}{lr} (\lambda_\perp/\lmhd)^{1/3}  \ , \ \ & {\rm for}\ 
\lambda_\parallel\ll \lpar \\
(\lambda_\parallel/\lmhd)^{1/2}
\ , 
\ \ &{\rm for}\  \lambda_\parallel\gg \lpar
\end{array} \right. \ .
\label{eq:3dspectrum}
\end{equation}
There is negligible additional power within an
eddy on parallel lengthscales smaller than $\lpar$, so 
for $\lambda_\parallel\ll\lpar(\lambda_\perp)$, 
$v_{\lambda_\perp,\lambda_\parallel}=v_{\lambda_\perp}$.  
For $\lambda_\parallel\gg\lpar(\lambda_\perp)$, 
the smallest eddy that contains both $\lambda_\perp$ and $\lambda_\parallel$
has a transverse lengthscale $\lambda_\perp'$ which
satisfies $\lpar(\lambda_\perp')=\lambda_\parallel$.  
The velocity fluctuation of this eddy is
obtained by solving this equation 
for $\lambda_\perp'$
(eq. [\ref{eq:spectrum2}]), and 
inserting this $\lambda_\perp'$ in equation (\ref{eq:spectrum1}).
Contours of the three-dimensional spectrum are plotted in Figure 
\ref{fig:contours}. Each contour represents eddies of a characteristic size.

\citet{mg01} give the three-dimensional spectrum in Fourier-space.
Since eddies that are separated by more than 
$\lpar$ are statistically independent, 
the power spectrum at a fixed transverse wavenumber $k_\perp$
is independent of the parallel wavenumber $k_\parallel$ in the
corresponding region of Fourier-space, i.e. where
$k_\parallel^{-1}\gtrsim\lpar(k_\perp^{-1})$.

The turbulent cascade is generally viewed as ``proceeding'' from larger eddies 
to smaller eddies as this is the direction of energy transfer. However, smaller 
eddies cascade many times in the time that a large eddy cascades. This is 
particularly important in turbulent mixing. Consider the evolution of two fluid 
elements whose initial separation is larger than the dissipation scale. On
cascade timescale $t_{\lambda_\perp}$, 
their transverse separation will random walk
a distance $\lambda_\perp$ as the result of the cascade of smaller eddies. 
Therefore, on a timescale comparable to an eddy's cascade time, the transverse 
locations of its component fluid elements---whose sizes may be considered to be
comparable to the dissipation scale---are completely randomized.  Moreover,
since mixing at the dissipation scale causes neighbouring fluid elements to be 
rapidly homogenized, transverse smoothing of the eddy occurs on the timescale 
that it cascades. Rapid transverse mixing in MHD turbulence is similar to the 
more familiar isotropic mixing in hydrodynamic turbulence.

\subsection{Passive Scalar Spectrum}
\label{sec:passivescalar}

A passive scalar, $\sigma$, satisfies the continuity equation, 
$(\partial/\partial t+\bld{v}\cdot \bld{\nabla})\sigma=0$, and does not affect 
the fluid's evolution. It could represent, for example, the concentration
of a contaminant.  We consider the spectrum of a passive scalar mixed
by the Alfv\'en wave cascade. These considerations are important
for our subsequent investigation of compressible turbulence. They are also
helpful for understanding the slow wave spectrum. We discuss the passive scalar 
spectrum both in the inertial range and also below the scale at which the 
Alfv\'en wave spectrum is cutoff.

\subsubsection{Passive Scalar Spectrum in the Inertial Range}

As we show in this subsection, 
the transverse spectrum of the passive scalar in the
inertial range is
\begin{equation}
\sigma_{\lambda_\perp}\propto \lambda_\perp^{1/3}
\label{eq:cspectrum1}  \ ,
\end{equation}
where $\lambda_\perp$ is the lengthscale transverse to the local mean
magnetic field and $\sigma_{\lambda_\perp}$ is
the r.m.s. fluctuation in the passive scalar
across $\lambda_\perp$.
The passive scalar parallel spectrum is the same as
the Alfv\'en wave parallel spectrum given in equation (\ref{eq:spectrum2}),
where $\lpar$ is now to be interpreted as
the lengthscale
parallel to the local mean magnetic field across which
the passive scalar fluctuation is $\sigma_{\lambda_\perp}$.

Mixing of the passive scalar is due to Alfv\'en waves. Slow wave mixing is 
negligible. This is because transverse velocity gradients are much larger than 
parallel ones in MHD turbulence. Thus Alfv\'en waves, whose velocity 
fluctuations are perpendicular to the magnetic field, are much more effective at 
mixing than slow waves, whose velocity fluctuations are nearly parallel to
the magnetic field. The transverse cascade arises from the shuffling of 
fieldlines as Alfv\'en waves propagate through the fluid.

The transverse spectrum (eq. [\ref{eq:cspectrum1}]) follows from
the Kolmogorov-like hypothesis that the cascade
rate of the ``energy'' in the scalar field is independent
of lengthscale; i.e. $\sigma_{\lambda_\perp}^2/t_{\lambda_\perp}$
is constant,
where $t_{\lambda_\perp}$ is the passive scalar cascade time,
which is assumed to be proportional to the cascade time of
Alfv\'en waves.  Comparing this
with the constancy of the kinetic energy cascade rate,
$v_{\lambda_\perp}^2/t_{\lambda_\perp}$, we conclude that
$\sigma_{\lambda_\perp}\propto v_{\lambda_\perp}$, which implies
equation (\ref{eq:cspectrum1}).
A similar argument holds for the cascade of a passive
scalar in hydrodynamic turbulence (e.g., Tennekes \& Lumley 1972).

The parallel cascade of a passive scalar is more subtle. It might appear that 
a passive scalar cannot cascade along fieldlines since, neglecting dissipation,
both the scalar and the magnetic field are frozen to the fluid, and thus the 
scalar must be frozen to fieldlines. In that case there certainly could not be a 
parallel cascade. If the scalar were injected on large scales, then fluctuations
with smaller wavelengths along the magnetic field would not be generated. 
However, dissipation cannot be neglected. It is an essential part of MHD 
turbulence, as it is of hydrodynamic turbulence.  For example, the description 
of turbulent mixing in \S \ref{subsec:eddies} depends crucially upon small scale 
dissipation.

Perhaps the best way to understand the parallel cascade is to consider
mixing on the transverse lengthscale $\lambda_\perp$ within two planes 
which are perpendicular to the local mean magnetic field, and which are 
separated by a parallel distance larger than $\lpar$. Velocity fluctuations 
within the two planes are statistically independent. This is evident because a 
bundle of fieldlines cannot be localized within a transverse distance 
$\lambda_\perp$ over a parallel separation greater than $\lpar$. Even a pair of 
fluid elements, one in each plane, that are initially on the same fieldline are 
mixed into two regions with different values of passive scalar concentration. 
It follows that the parallel cascade of the passive scalar also obeys equation 
(\ref{eq:spectrum2}).  

\subsubsection{Passive Scalar Spectrum Below Alfv\'en Wave Cut-Off}
\label{sec:cutoff}

A passive scalar cascade may extend below the transverse scale at which the MHD 
cascade is cut-off. Mixing on these scales is driven by fluid motions at 
$\lambda_{\rm cutoff}$ which results in a scale-independent mixing time equal to 
the cascade time at $\lambda_{\rm cutoff}$. This yields
\begin{equation}
\sigma_{\lambda_\perp}= {\rm constant}  \ , \ \ \
\lambda_\perp<\lambda_{\rm cutoff} \ . \label{eq:cspectrum3}
\end{equation}
A similar argument applies in hydrodynamic turbulence.
\citet{tl72} call this regime in hydrodynamic turbulence
the ``viscous-convective subrange''.

\subsection{Slow Wave Spectrum}
\label{sec:slowwavespectrum}

The slow wave spectrum is the same as that of the Alfv\'en waves. It is given by 
equations (\ref{eq:spectrum1}) and (\ref{eq:spectrum2}), with 
$v_{\lambda_\perp}$ referring to the velocity fluctuations of the slow waves.
This is a consequence of the similar kinematics of 
slow waves and shear Alfv\'en 
waves and the fact that both are cascaded by shear Alfv\'en waves.

Slow waves obey the same linear wave equation as Alfv\'en waves, and to lowest 
nonlinear order they travel up and down the local mean magnetic field lines at 
the Alfv\'en speed just as Alfv\'en waves do. However, the dynamics of the MHD 
cascade is controlled by the Alfv\'en waves \citep{gs97, mg01} because their
velocity and magnetic field fluctuations are perpendicular to the local mean 
magnetic field, whereas those of the slow waves are nearly parallel to it.  
Since perpendicular gradients are much larger than parallel ones in 
the MHD cascade, Alfv\'en waves are much more effective at mixing than are slow 
waves. Hence, Alfv\'en waves cascade both themselves and slow waves, whereas 
slow waves cascade neither.\footnote{We are assuming that Alfv\'en and slow 
waves have comparable strengths at a given lengthscale.}

The transverse mixing of the slow waves by Alfv\'en waves is analogous to the 
mixing of a passive scalar.  As discussed in \S \ref{sec:passivescalar}, a 
passive scalar assumes the same inertial range spectrum as that of the velocity 
field which is responsible for its mixing.  Thus, equation (\ref{eq:spectrum1})
is also applicable to the velocity fluctuations of the slow waves.

Similarly, the parallel cascade of slow waves is analogous to the parallel 
cascade of a passive scalar. Since Alfv\'en waves
cascade in the time they move a distance $\lpar$, slow waves separated by this 
distance are independently mixed.
Thus, equation (\ref{eq:spectrum2}) also applies to slow waves.
There is, however, a conceptual difference between the parallel cascades
of the passive scalar and of the slow mode.  In the absence of dissipation a passive 
scalar is frozen to fieldlines, whereas slow mode wave-packets travel along 
them at the Alfv\'en speed. A passive scalar has a parallel cascade because
Alfv\'enic fluctuations are statistically independent within two transverse 
planes {\it frozen in the fluid} and separated by $\lpar$. The parallel cascade 
of slow waves occurs because two 
transverse planes separated by $\lpar$ and {\it 
travelling at the Alfv\'en speed in the same direction} experience uncorrelated  
sequences of distortions suffered as a result of interactions with oppositely 
directed  Alfv\'en waves. 
Nevertheless, these two requirements are both satisfied in the
MHD cascade when the transverse planes are separated by a distance
greater than $\lpar$, and so the passive scalar and the slow mode
have the same parallel spectrum.
Whereas passive scalar mixing is due to eddies, slow wave mixing
may be thought of as due to ``travelling eddies'', i.e. eddies which
travel up and down the magnetic field at the Alfv\'en speed.

\subsection{Numerical Simulations}

Numerical simulations offer some support for the above description of 
incompressible MHD 
turbulence. Those by \citet{cv00b} support both equations (\ref{eq:spectrum1}) 
and (\ref{eq:spectrum2}), and those by \citet{mb00} support equation 
(\ref{eq:spectrum1}). However, although the simulations of \citet{mg01} support 
equation (\ref{eq:spectrum2}), they yield $v_{\lambda_\perp}\propto 
\lambda_\perp^{1/4}$ instead of equation (\ref{eq:spectrum1}). Because the 
simulations of \citet{mg01} are stirred highly anisotropically, whereas those of 
\citet{cv00b} and \citet{mb00} are stirred isotropically, it is not clear
whether these disparate results conflict. \citet{mg01} speculate that the 
discrepancy between their spectrum and the scaling prediction of \citet{gs95} 
results from intermittency. In any case, we expect that the physical picture of 
a critically balanced cascade, which underlies Goldreich \& Sridhar's 
description of MHD turbulence, remains valid. Even if the spectrum is
proportional to $\lambda_\perp^{1/4}$, we expect that the results of this 
investigation---which assumes a spectrum proportional to  
$\lambda_\perp^{1/3}$---would
not be significantly altered.

The simulations of \citet{mg01} confirm that a passive scalar 
has the same transverse spectrum as that of Alfv\'en waves, although both are 
proportional to $\lambda_\perp^{1/4}$. They also indicate that the parallel
cascade of a passive scalar conforms to equation (\ref{eq:spectrum2}).

\citet{mg01} present results from simulations of the interaction between 
oppositely directed slow and Alfv\'en waves;  the slow waves cascade
whereas the Alfv\'en waves do not.
They also compute the spectrum of slow waves in a 
simulation of MHD turbulence and find that both its transverse and longitudinal 
behavior matches that of the Alfv\'en wave spectrum.

\subsubsection{Kolmogorov Constants}
\label{sec:kolmogorov}

Scaling arguments do not yield values for the  ``Kolmogorov constants'', the 
order-unity multiplicative constants of the spectrum. However, they can be
obtained from simulations. We define them such that equation 
(\ref{eq:spectrum1}) remains valid, i.e. we take $\lmhd$ to be the separation at 
which the r.m.s. velocity difference is equal to, or extrapolates to, $v_A$. Two 
Kolmogorov constants, $M_\parallel$ and $M_t$, are needed in this paper:
\begin{eqnarray}
M_\parallel&\equiv& {v_A\over v_{\lambda_\perp}}
{\lambda_\perp\over \lpar}
\Rightarrow \lpar=
M_\parallel^{-1}\lambda_\perp^{2/3}\lmhd^{1/3} \label{eq:kol1} \\
M_t&\equiv&{v_A t_{\lambda_\perp}\over \lpar} \label{eq:kol2} \ .
\end{eqnarray}
In these definitions, $\lambda_\perp$ and $\lpar$ are the inverses 
of the wavenumbers transverse and parallel to the local mean magnetic field,
and $t_{\lambda_\perp}$ is the cascade time of waves with transverse wavenumber 
$\lambda_\perp^{-1}$. From the numerical simulations of \citet{mg01}, 
\begin{equation}
M_\parallel\simeq 3.4 \ \ \ {\rm and} \ \
M_t\simeq 1.4 \ .
\label{eq:kolmogorov}
\end{equation}
Because these simulations yield a transverse spectrum which is 
proportional to $\lambda_\perp^{1/4}$ instead of $\lambda_\perp^{1/3}$,
the resulting ``Kolmogorov constants'' are not truly constant.

\section{COMPRESSIBLE TURBULENCE: OVERVIEW}

Our primary concern is interstellar scintillation, 
which is affected by electron 
density fluctuations on very small scales, typically $10^8-10^{10}$ cm for 
diffractive scintillation. In the remainder of this paper
we calculate the spectrum of density fluctuations which results
from compressible turbulence in magnetized plasmas with parameters
appropriate to the interstellar medium.
Throughout, we consider plasmas that have more ions than neutrals, 
and that have 1 $\lesssim\beta<\infty$,
where $\beta$ is the ratio of the thermal pressure to the mean magnetic
pressure: $\beta\equiv 8\pi p/B^2=2c_T^2/v_A^2$.
Here, $c_T\equiv (p/\rho)^{1/2}$ is the isothermal sound speed,
$v_A$ is the Alfv\'en speed, $p$ is the thermal pressure,
$B$ is the magnetic field strength, and $\rho$ is the mass 
density.\footnote{We briefly discuss plasmas with $\beta<1$ in 
\S \ref{sec:betalt1}.}
The incompressible limit corresponds to $\beta=\infty$.

On the lengthscales that we consider,
compressible MHD is a good approximation
for the dynamics of the ionized interstellar medium.
Kinetic effects, where important, may be accounted for
by simple modifications to the MHD equations.
Therefore, we turn our attention to turbulence in compressible MHD.

The turbulent velocity spectrum in compressible MHD is
approximately the same as the turbulent
velocity spectrum in incompressible MHD,
because the Alfv\'en mode remains incompressible in a compressible medium,
and the slow mode is only slightly compressible.
Thus, the velocity spectrum for both of these modes is given
by equations (\ref{eq:spectrum1}) and (\ref{eq:spectrum2}).
The Appendix summarizes
the properties of the relevant modes in compressible MHD.

There are two additional 
modes in compressible MHD which are not present in incompressible
MHD.  One of these is the fast mode.  However, 
as long as $\beta\gtrsim$ a few, the fast mode is essentially a 
sound wave.
Its phase speed is larger than the phase speed of either the Alfv\'en
mode or the slow mode, and so the fast mode does not couple to them.
Thus, we ignore the fast mode.  An analogous approximation is
often made in subsonic hydrodynamic turbulence, where sound waves have little
influence on the turbulent cascade.

The second mode that is present only in compressible MHD is the
entropy mode.  This is a zero-frequency mode
with unperturbed pressure, and with density perturbation offset by
temperature perturbation (see Appendix).
The entropy mode exists when the turbulent
motions are adiabatic; the entropy mode is then the only source
of entropy fluctuations.
In hydrodynamics,  simple
scaling arguments show that the entropy mode does not affect the dynamics
of the fluid---i.e. the fluid obeys the incompressible equations
of motion---provided that fluid motions are subsonic and
that density fluctuations are smaller than the mean density;
these arguments carry over directly to MHD 
(e.g. Higdon 1986, and references therein).
Furthermore, since the entropy of any fluid element
is conserved in the inertial range of adiabatic turbulence,
the mixing of the entropy mode is
identical to the mixing of a passive scalar
(see eq. [\ref{eq:cspectrum1}]), yielding the transverse spectrum
\begin{equation}
s_{\lambda_\perp}\propto \lambda_\perp^{1/3} \ ,
\end{equation}
where $s_{\lambda_\perp}$ is the r.m.s. entropy fluctuation across
$\lambda_{\perp}$.  The parallel spectrum is given by
equation (\ref{eq:spectrum2}).
Analogously, in hydrodynamic turbulence the effect of the entropy
mode on turbulent motions can often be neglected, and
the entropy mode is mixed
as a passive scalar \citep{my71}.

Based on the above discussion, there are two sources of density
fluctuations on small scales: the slow mode and the entropy mode.  Since the 
slow mode density perturbation is 
proportional to its velocity perturbation, and since
the entropy mode density perturbation is proportional to its entropy 
perturbation, both yield a Kolmogorov spectrum of density perturbations:
\begin{equation}
n_{\lambda_\perp} \propto \lambda_\perp^{1/3}.
\label{eq:density}
\end{equation}

In the remainder of this paper, we investigate the density spectrum
in more detail. 
We calculate the density spectrum on diffractive lengthscales
for given values of the number density, $n$, and 
outer scale, $\lout$.
Typically, in regions of the interstellar medium
which are relevant for scintillation,
$1\ {\rm cm}^{-3}<n<100$ cm$^{-3}$.
Values of $\lout$ are more uncertain, though it is likely that
$\lout$ is typically within a few orders of magnitude of 1 pc. It is also
plausible that the value of $\lout$ is related to that of $n$.
In paper II, we discuss in much more detail the values of these
parameters in turbulent regions of the ionized interstellar medium,
e.g. in supernova shocks, HII regions, and stellar winds.

A naive guess for the resulting density fluctuation at the
lengthscale $\lambda_\perp$ is
$n_{\lambda_\perp}=n(\lambda_\perp/\lout)^{1/3}$, which we refer to as
the fiducial spectrum.
However, there are a number of physical effects which suppress the small-scale 
density spectrum in the interstellar medium relative to the fiducial spectrum.

When considering scintillation observations, it is the
transverse---not parallel---lengthscale which is
relevant.
Each line of sight crosses
regions with different orientations of the
local mean magnetic field.
The observational effects of the parallel lengthscale are washed out
if the orientation varies by an angle
greater than a tiny ratio:  
$\lambda_\perp/\lpar$ at the diffractive scale.
Since the variation in angle due to eddies on scales
larger than the diffrative scale is
$\lambda_\perp/\lpar\propto\lambda_\perp^{1/3}$,
it increases with scale, so
these
eddies render the parallel lengthscale unobservable.
As a result, we frequently refer to the
transverse lengthscale as, simply, the lengthscale.

Because of the multitude of special lengthscales involved,
we organize the discussion by decreasing lengthscale.
We begin at the outer scale, and proceed down to the smallest scales
relevant to interstellar scintillation, while considering the effects of the
Alfv\'en mode, the slow mode, and
the entropy mode  simultaneously.
Since most of the relevant lengthscales in the interstellar medium have
a similar dependence on the background density---they
decrease with increasing density---the
ordering of lengthscales is fairly universal.
The lengthscales which we consider are the outer scale,
the MHD scale, the cooling scale, the collisionless scale of the neutrals,
the collisionless scale of the ions, and
the proton gyro scale.
We conclude with a summary of the most important effects.
Table \ref{tab:lengths} lists the lengthscales that are used
in this paper.  

\section{THE OUTER SCALE AND THE MHD SCALE}

As a model for the excitation of the turbulence, we consider
plasma which is stirred on an outer scale $\lout$ with velocity 
fluctuations that are of order the sound speed: $v_{\lout}\sim c_s$.
The generalization from this case of Mach 1 turbulence
to subsonic turbulence with arbitrary Mach number 
${\cal{M}}<1$ is trivial: 
$\lout$ would be interpreted as an effective outer scale at which velocity
fluctuations extrapolate to $c_s$.  However, we focus on the Mach 1
case because it is probably the most relevant for interstellar scintillation.

If, initially, the strength of the magnetic field is negligible, then
random fieldline stretching amplifies the mean field \citep{cv00a}.
It is uncertain both how quickly the magnetic field
is amplified, and whether its energy density is amplified until it approaches 
equipartition with the turbulent kinetic energy density.  If it does reach 
equipartition within a few outer-scale crossing times, then $\beta\sim 1$ would 
be appropriate for Mach 1 turbulence. 
Recall that $\beta$ is the ratio of thermal pressure
to magnetic pressure, or equivalently, $\beta=2 c_T^2/v_A^2$.
We generally assume that the mean magnetic field can
approach equipartition with the gas pressure, so we think of $\beta$
as close to but a little larger than
unity, although we leave its exact value unspecified.

Provided $\beta$ is larger than unity and ${\cal{M}}\sim 1$ on scale $\lout$,
the turbulent kinetic energy dominates the mean magnetic energy on scales just
below $\lout$. Thus the cascade is hydrodynamic on these lengthscales, and the 
velocity fluctuations are given by Kolmogorov's isotropic scaling:
$v_{\lambda_\perp}\sim c_s (\lambda_\perp/\lout)^{1/3}$. 

The kinetic energy density decreases towards smaller scales. At a 
critical scale---which we
denote $\lmhd$---the kinetic energy density is 
sufficiently small that it is comparable to the
mean magnetic energy density: $v_{\lmhd}^2\sim v_A^2$, which implies that
\begin{equation}
\lmhd\sim\lout\beta^{-3/2} \ .
\label{eq:lambdamhd}
\end{equation}
Below this scale, the turbulent kinetic energy density is smaller than the 
magnetic energy density, and the theory of compressible MHD turbulence is 
applicable. Note that the effects of large-scale velocity fields
can be neglected when considering
fluctuations on smaller scales, because a large-scale velocity field
can be eliminated by a Galilean transformation.
However, since large-scale magnetic fields cannot be transformed
away, they affect the dynamics of small-scale eddies. 

Hydrodynamic turbulent motions on scales slightly larger than $\lmhd$ have
speeds $\sim v_A$. Thus they couple to, and efficiently excite, Alfv\'en waves 
and slow waves whose phase velocities $\omega/k\approx v_A(k_z/k)$ are of 
comparable magnitude. Provided $\beta>1$ and ${\cal M}\leq 1$, fast waves, which 
have $\omega/k\sim c_s > v_A$, are weakly excited at $\lmhd$. In what follows, 
we neglect fast waves. 

Alfv\'en waves excited at $\lmhd$ cascade to smaller scales,
and these cascading Alfv\'en waves in turn cascade slow waves.
The spectra of both of these cascades are given by equations
(\ref{eq:spectrum1}) and (\ref{eq:spectrum2}), so that
\begin{equation}
v_{\lambda_\perp}\sim c_s \Big({\lambda_\perp\over\lout}\Big)^{1/3}
\label{eq:vspec}
\end{equation}
for both modes, even on scales smaller than $\lmhd$.

We have not yet discussed density fluctuations on the
scales which have been considered in this section.
It is more convenient to do so in the following section.

\section{THE COOLING SCALE}

On the lengthscales in the interstellar medium that we have considered thus 
far, the radiative cooling time is shorter than the eddy turnover 
time.\footnote{For definiteness, we consider plasma that is thermally stable;
deviations of the temperature from its equilibrium value decay
in a cooling time, $t_{\rm cool}$.} Consequently, in photoionized regions, 
turbulence on these lengthscales is expected to be isothermal (Sridhar 1998; 
Goldreich 1998; Higdon \& Conley 1998).
This has only a marginal effect
on the turbulent dynamics described in the previous section,
because isothermal Alfv\'en waves are identical to adiabatic
Alfv\'en waves, and isothermal slow waves are only slightly
different from adiabatic slow waves.  However,
there is no entropy mode in isothermal
turbulence.  As a consequence, small-scale density fluctuations
may be significantly suppressed.
There are two possible solutions to
this ``cooling catastrophe'': either (i) the outer scale is extremely
small, small enough that the turbulence at the outer scale
is nearly adiabatic; or (ii) there are significant density fluctuations
associated with
the slow mode.  However, in the latter case,
the mean magnetic
field must be amplified almost to equipartition with the gas pressure,
so that $\beta\sim 1$.
Either of these two solutions would place stringent constraints on
the nature of the turbulence which is responsible for observed
density fluctuations.  In the following, we consider the cooling scale in
more detail.

There is a critical scale for the turbulence, which we
call the ``cooling scale'', $\lcool$.  Above
this scale the turbulence is isothermal, and below it
the turbulence is adiabatic.
We assume that $\lcool<\lmhd$ throughout this paper,
except in \S \ref{sec:coolinglargebeta} where we consider
the case in which this inequality is reversed. The cooling scale is where
the  eddy turnover time, $\lambda_\perp/v_{\lambda_\perp}$, is equal to
the cooling time, $t_{\rm cool}$; i.e., with $v_{\lambda_\perp}\sim
c_s(\lambda_\perp/\lout)^{1/3}$,
\begin{equation}
{\lcool\over \lout }
\sim
\Big( {c_s t_{\rm cool}\over \lout} \Big)^{3/2}  \ .
\label{eq:coolingscale}
\end{equation}

As we will discuss in detail in paper II, most plausible astrophysical sources 
of interstellar scintillation have $c_st_{\rm cool} \lesssim \lout$, and thus
$\lcool\lesssim\lout$. In general, the kinetic power per unit mass which is
dissipated in Mach 1 turbulence is $\sim c_s^2/t_{\lout}\sim
c_s^3/\lout$, where $t_{\lout}$ is the eddy
turnover time at the outer scale. The thermal power per unit mass
which is required to keep the gas hot is $\sim c_s^2/t_{\rm cool}$.
Hence, to avoid the conclusion that the cooling scale is
smaller than the outer scale, one must require that more energy
go into turbulent motions than into heat.  As we will see in paper
II, this requirement is difficult to satisfy.
Nonetheless, in isothermal shocks, and in plasmas which are thermally
unstable, the two powers are comparable, and the cooling scale is
comparable to the outer scale.

As an example, we consider an HII region.
Photoionized plasma is thermally stable. Heating is
primarily due to photoionizing photons, and cooling is primarily
due to electron impact excitation of metal line transitions (e.g., Spitzer 
1978).\footnote{
In this case, there is another lengthscale,
the ``photoionization scale'', which we do not consider because
its effects are unimportant for scintillations.  This is the scale
at which the eddy turnover time is comparable to the
recombination time.  It is slightly larger than the cooling
scale because the recombination time is approximately five times larger than the
cooling time \citep{spitzer78}.
On scales larger than the photoionization scale,
the turbulence is in photoionization equilibrium, whereas on scales
smaller than this, the ionization fraction of a fluid element is conserved.
Nonetheless, the temperature of the gas is only weakly dependent on
ionization fraction: metal-line cooling by free electrons is exponentially
dependent on temperature, and thus acts as a thermostat which is
only slightly affected by the ion or neutral density.
Thus, the photoionization lengthscale
does not play an important role in the density spectrum.}
A characteristic temperature for plasma photoionized by hot stars is
$T\sim$ 8,000K, which implies that the speed of sound is
$c_s\sim 10 $ km/sec.
The cooling time is
$t_{\rm cool}\sim 20,000 (\cm^{-3}/n)$ years,
where $n$ is the number density of electrons.
Therefore,
\begin{equation}
c_s t_{\rm cool}\sim
0.2
\Big({{\rm cm}^{-3}\over n}\Big)
\ \  {\rm pc} \ .
\label{eq:cstcool}
\end{equation}
A plausible value for $\lout$ might be the radius
an HII region (the ``Str\"omgren radius''), which is
$\sim 70 (\cm^{-3}/n)^{2/3}$ pc \citep{spitzer78}.
Therefore, with typical values of $n$ in
HII regions ($1\ {\rm cm}^{-3}<n<100$ cm$^{-3}$), $\lout$
is significantly larger than $c_st_{\rm cool}$, and so the
cooling scale is significantly smaller than the outer scale.

The fact that the turbulence is isothermal on large scales
has important implications.
Had cooling been ignored, i.e. had it been implicitly assumed
that the cooling time is infinite, then
one would have calculated the density
spectrum as follows:  there should be density fluctuations of
order unity on the outer scale, implying excitation of
entropy modes on the
outer scale which are passively mixed by the Alfv\'enic
turbulence to small scales. This would yield the fiducial density
spectrum, $n_{\lambda_\perp}\simeq n (\lambda_\perp/\lout)^{1/3}$.

However, if the turbulence is isothermal on large scales, the
above calculation is incorrect:  the entropy mode does not exist
under isothermal conditions.
In the following, we consider separately two cases:
first the low-$\beta$ case ($\beta<\lout/c_st_{\rm cool}$),
and then the high-$\beta$ case ($\beta>\lout/c_st_{\rm cool}$).
In each case, we show that the small-scale density fluctuations
are substantially smaller than those predicted by the fiducial spectrum unless
a relatively extreme condition holds: either $\beta\sim 1$ 
or $\lout\sim c_s t_{\rm cool}$.

\subsection{Density Fluctuations Below the Cooling Scale:
$1<\beta<\lout/c_st_{\rm cool}$}
\label{sec:coolingsmallbeta}

With regards to interstellar scintillation, the crucial difference between 
high-$\beta$ and low-$\beta$ turbulence 
lies in the compressibility 
of the slow mode.  Since the sum of thermal pressure and magnetic pressure 
vanishes for the slow mode, the mode's density perturbation satisfies 
$n_{\lambda_\perp}/n\sim p_{\lambda_\perp}/p\sim
\beta^{-1}B_{\lambda_\perp}/B\sim \beta^{-1} v_{\lambda_\perp}/v_A\sim
\beta^{-1/2}v_{\lambda_\perp}/c_s$ (see Appendix).
So, assuming that the energy in slow waves is comparable to that in Alfv\'en 
waves,
\begin{equation}
{n_{\lambda_\perp}\over n}\sim
{1\over \sqrt{\beta}}
\Big( {\lambda_\perp\over \lout} \Big )^{1/3} \
\ \ \ \ \ \lambda_\perp<\lmhd \ \ ,
\label{eq:slowdens}
\end{equation}
from the slow mode, both above and below the cooling scale.
Thus, slow waves can produce density fluctuations
which are not much smaller than the fiducial spectrum if $\beta$ is not
much larger than unity. Comparing with equation (\ref{eq:bigbeta}) below,
the contribution of the slow mode to the density fluctuations
at $\lcool$
exceeds the contribution from Reynolds stresses provided
$\beta<\lout/c_st_{\rm cool}$.

The compressibility of the slow mode also yields density fluctuations
associated with the entropy mode.  The reason is as follows. Entropy 
fluctuations are associated with isothermal compressible waves. Those
due to isothermal slow modes are passively mixed by Alfv\'en waves. Mixing due 
to Alfv\'en waves with wavelengths smaller than the cooling scale takes place in 
less than the cooling time. It produces a spectrum of entropy modes for 
$\lambda_\perp<\lcool$ that gives rise to a density spectrum similar to the one given 
in equation (\ref{eq:slowdens}).  
Nonetheless, there is some damping of entropy fluctuations when they are
mixed from scales larger than the cooling scale to scales smaller
than the cooling scale.  We expect that the damping is of
order unity.
In this paper, however, 
we do not quantify the amount of damping more precisely, because it is
a difficult calculation; it is more difficult than the damping
of the slow mode, which we calculate in the following section.

\subsubsection{Slow Mode Damping at the Cooling Scale}
\label{sec:slowdamp}

Although there is negligible damping of the slow mode
on lengthscales which are either
much larger or much smaller than the cooling scale, there
is some damping on lengthscales which are comparable to the
cooling scale.
As we show in this section,
the damping of the slow wave cascade is small, because the slow
mode is not very compressible.
In the limit that $\beta\rightarrow\infty$, the slow mode is incompressible,
and the damping disappears.  In the following, we calculate the damping to
first order in $1/\beta$.

First, we calculate the damping rate of a slow wave of a fixed
wavenumber.  From the Appendix,
the dispersion relation of the slow mode,
in the limit that $k_\parallel/k\ll 1$, is
\begin{equation}
\omega=v_A k_\parallel  {\tilde{c}\over\sqrt{\tilde{c}^2+v_A^2}}  \ .
\end{equation}
The notation which we use in this section is the same as the notation
which we define in the Appendix,
except that here we label the axis
parallel to the mean magnetic field by $\parallel$ instead of by $z$.
Note that we define $\tilde{c}^2\equiv p'/\rho'$,
where primes denote Eulerian perturbations.
The linearized equation of energy conservation is
\begin{equation}
i\omega Ts'={T'\over {t_{\rm cool}}} \ ,
\end{equation}
where $s'$ is the perturbed entropy per particle.
Combining this equation with the following
monatomic ideal gas relations
\begin{equation}
{p'\over p}={\rho'\over \rho}+{T'\over T}=
{2\over 3}s'+{5\over 3}{\rho'\over \rho}  \ ,
\end{equation}
yields
\begin{equation}
\tilde{c}^2\equiv {p'\over \rho'}={c_T^2}
{ {2-i5\omega t_{\rm cool}} \over
  {2-i3\omega t_{\rm cool}} } \ ,
\label{eq:tildec}
\end{equation}
where $c_T^2\equiv p/\rho=\beta v_A^2/2$ is the square of the isothermal
sound speed.   On large scales, $\omega t_{\rm cool}\ll 1$, and
$\tilde{c}$ is the isothermal sound speed; on small scales,
$\omega t_{\rm cool}\gg 1$, and $\tilde{c}$ is the adiabatic sound speed.
The damping rate is
obtained by substituting the above $\tilde{c}$
(eq. [\ref{eq:tildec}])
into the dispersion
relation, which gives, to lowest order in $1/\beta$,
\begin{equation}
{\omega_i \over v_A k_\parallel }
=
{-v_A k_\parallel t_{\rm cool} \over
 \beta [1+(5v_A k_\parallel t_{\rm cool}/2)^2] } \ , \label{eq:imcool}
\end{equation}
where $\omega_i$ is the  imaginary part of $\omega$.
Only on lengthscales where $v_Ak_\parallel t_{\rm cool}\sim 1$ is
this ratio non-negligible.

To obtain the total
decrement in the amplitude of the slow mode through
the cooling scale, we solve a kinetic equation obtained by balancing
the slow mode $k$-space energy flux with the loss-rate of
slow mode $k$-space
energy density due to damping; i.e.
\begin{equation}
{d\over {dk}} {v_{\lambda_\perp}^2\over t_{\lambda_\perp}}
=2\omega_i {v_{\lambda_\perp}^2 \over k} \ ,
\end{equation}
where $\lambda_\perp^{-1}\equiv k_\perp$ is the transverse
wavenumber,
$t_{\lambda_\perp}$ is the cascade time of a slow wave with
this wavenumber, and $v_{\lambda_\perp}$ is the velocity perturbation
of the slow mode; since $k_\perp\simeq k$, we drop the $\perp$ 
subscript on $k$.
We rewrite the kinetic equation as follows:
\begin{equation}
{d\over d\ln k}
\ln {v_{\lambda_\perp}^2\over t_{\lambda_\perp}}
= 2\omega_i t_{\lambda_\perp}
=2M_t
{\omega_i\over v_A k_\parallel}  \ ,
\label{eq:coolcas}
\end{equation}
where  $M_t\equiv v_A k_\parallel t_{\lambda_\perp}\simeq 1.4$
is a Kolmogorov constant (see \S \ref{sec:kolmogorov}).
Substituting equation (\ref{eq:imcool}) into equation
(\ref{eq:coolcas}) and integrating, using
$k_\parallel\propto k^{2/3}$
(eq. [\ref{eq:spectrum2}]) and $t_{\lambda_\perp}\propto k^{-2/3}$,
yields the net damping through the cooling scale:
\begin{eqnarray}
{[v_{\lambda_\perp}/\lambda_\perp^{1/3}]\vert_{{\lambda_\perp}\ll\lcool}
\over
[v_{\lambda_\perp}/\lambda_\perp^{1/3}]\vert_{{\lambda_\perp}\gg\lcool}}
&=&
\exp\Big({
-{M_t\over \beta}\int_0^\infty
{v_A k_\parallel t_{\rm cool} \over
 1+(5v_A k_\parallel t_{\rm cool}/2)^2 }
{dk\over k}
}\Big) \nonumber \\
&=&
\exp\Big(-{3\pi M_t \over 10\beta}\Big) \nonumber \\
&\simeq&
1-1.3/\beta \ ,
\end{eqnarray}
to lowest order in $1/\beta$.  Thus for $\beta$ slightly larger
than unity, slow mode damping can be ignored.

\subsection{Density Fluctuations Below the Cooling Scale:
$\beta>\lout/c_st_{\rm cool}>1$}
\label{sec:coolinglargebeta}

In this subsection only, we assume that $\beta>\lout/c_st_{\rm cool}$, which
implies $\lmhd<\lcool$. In this case, the isothermal hydrodynamic turbulent 
cascade extends from the outer scale down to the cooling scale. Because the 
entropy mode does not exist under isothermal conditions and hydrodynamic 
turbulence is incompressible to order $v_{\lambda_\perp}/c_s$, to this order 
there are no density fluctuations on lengthscales larger than
$\lcool$. However, pressure fluctuations due to Reynolds stresses, 
$p_{\lambda_\perp} 
\sim\rho v_{\lambda_\perp}^2$, create second order density 
and entropy fluctuations,
\begin{equation}
{n_{\lambda_\perp} \over n}={p_{\lambda_\perp}\over p}=s_{\lambda_\perp}
\sim \Big({v_{\lambda_\perp}\over c_s}\Big)^2 \ , \ \ \
\lcool < \lambda_\perp < \lout \ .
\end{equation}
These entropy fluctuations couple to the entropy mode at the
cooling scale.  On scales below the cooling scale, the entropy mode is mixed as 
a passive scalar yielding
\begin{eqnarray}
{n_{\lambda_\perp}\over n}&\sim& {v_{\lcool}^2\over c_s^2}
\Big( {\lambda_\perp\over \lcool} \Big )^{1/3} \nonumber \\
&\sim& \Big( {\lcool\over \lout} \Big)^{2/3}
\Big( {\lambda_\perp\over \lcool} \Big )^{1/3} \nonumber \\ 
&\sim&
\Big({c_s t_{\rm cool} \over \lout}\Big)^{1/2}
\Big( {\lambda_\perp\over \lout} \Big )^{1/3} \ ,
\ \ \lambda_\perp < \lcool \ .
\label{eq:bigbeta}
\end{eqnarray}
Hydrodynamic turbulence mixes the entropy mode down to
$\lmhd$.  At this scale, the hydrodynamic motions
turn into Alfv\'en and slow waves, so the entropy mode
continues to be mixed below $\lmhd$, and equation
(\ref{eq:bigbeta}) remains valid.
Comparing with equation (\ref{eq:slowdens}), 
the density fluctuations associated with the
entropy mode exceed those associated with 
the slow mode when 
$\beta>\lout/c_st_{\rm cool}$.

The entropy mode's density spectrum (eq. [\ref{eq:bigbeta}])
is smaller
than the fiducial spectrum by the small factor
$({c_s t_{\rm cool}/\lout})^{1/2}$. As will be seen
in paper II, it
yields density fluctuations that are too small to explain much
of the observed interstellar scintillation for most plausible
values of $\lout$, such as those given by Str\"omgren radii
of HII regions.  There are two alternatives;
either $\lout$ is not much larger than
$c_s t_{\rm cool}$, or
$\beta<\lout/c_st_{\rm cool}$. 

\section{THE COLLISIONLESS SCALE OF THE NEUTRALS}

In the remainder of this paper, we consider lengthscales
which are comparable to, or smaller than, the mean free
paths of the various particles; thus, kinetic effects
must be considered.

In this section, we calculate the damping of the turbulence by neutrals.
We assume that the neutral density
is smaller than the electron density, i.e. $n_N\lesssim n$,
and that most of the ions are protons.
The neutral particles which we consider are hydrogen
atoms and helium atoms.  The relevant lengthscales are the
hydrogen and helium mean free paths for collisions with protons:
${\rm L}_{{\rm H}}=5\times 10^{13}(\cm^{-3}/n)$ cm
and
${\rm L}_{{\rm He}}=1.5\times 10^{15}(\cm^{-3}/n)$ cm
\citep{banks66}. The hydrogen mean free path is significantly
smaller than the helium mean free path because its
collisions with protons are due to resonant charge
exchange.

There are two main results for this section.
\begin{itemize}
\item{(i)} Only if the neutral fraction
is sufficiently small (eq. [\ref{eq:neutralfrac}], below) can the
cascade survive on lengthscales smaller than ${\rm L}_{{\rm H}}$.
If the neutral fraction does not satisfy this condition, then all three
modes---Alfv\'en, slow, and entropy---are damped at the same lengthscale.
\item{(ii)} Regions where the cascade does not survive contribute to an
excess of density fluctuations on large scales. This might
explain observations which detect an excess of power
in large-scale density fluctuations.
\end{itemize}

The organization of the calculation is as follows:  first,
the change in wave frequency due to neutrals is calculated.
Second, we consider the effect of the frequency change on
the cascade.
Third, we consider the cases in which the neutrals are
hydrogen atoms and in which they are helium atoms.
Although helium atoms have a larger mean free path, we
consider them after neutral hydrogen atoms because they
are of lesser importance.
Finally, we consider regions where neutral damping 
terminates the cascade.

A similar, though less detailed
calculation is performed by \citet{gs95}
for the case in which the neutrals are hydrogen atoms.

\subsection{Frequency Change}

Consider an incompressible wave, either an Alfv\'en wave or
a slow wave in the incompressible limit. Although the slow wave
is, in fact, slightly compressible, this does not have a large
effect on the final result.  We calculate
the frequency change of an incompressible wave with fixed $\bld{k}$:
$k_\parallel$ along the mean magnetic field and $k_\perp$
transverse to the mean magnetic field, where 
$k_\parallel\ll k_\perp\simeq k$.

We define $\bld{v}$ to be the mean velocity of the protons averaged over the 
proton distribution function. Thus $\bld{v}$
satisfies the equation of motion for the Alfv\'en wave or for the
slow wave derived in the Appendix.
We then calculate the force that these protons exert on the neutrals.
Since this force is the same as the force of the neutrals
on the protons,
inserting it into the equation of motion for the protons yields
the change in frequency.

We denote the perturbed distribution function for the neutrals,
i.e. perturbed to first-order in $\bld{v}_N$, after
being Fourier-transformed in space and time, 
by $\delta\!f_N(\bld{k},\omega,\bld{v}_N)$, where $\bld{v}_N$ is the
velocity of the neutrals.  The evolution of the neutrals is determined
by the Boltzmann equation:
\begin{equation}
-i\omega \,\delta\!f_N + i(\bld{k}\cdot\bld{v}_N)\delta\!f_N=\cal{C} \ ,
\end{equation}
where $\cal{C}$ is the Fourier-transformed, perturbed, collision integral 
of the neutrals with
the protons.  Neutral-neutral collisions are negligible relative to
neutral-proton collisions.
The collision integral is simplified by making the
approximation that, on the timescale that a neutral collides with
protons, $\delta\!f_N$ is driven
towards a Maxwellian with mean velocity $\bld{v}$:
\begin{equation}
{\cal{C}}=\nu_{N,p}\bigg[
{ (\bld{v}\cdot\bld{v}_N) \over T/m_N}
{n_N \exp({-m_Nv_N^2/2T})\over (2\pi T/m_N)^{3/2}}
-\delta\!f_N\bigg] \ ,
\end{equation}
where the first term in the square brackets is generated by expanding 
a Maxwellian distribution function with mean velocity $\bld{v}$ to 
linear order in $\bld{v}$ and retaining only the perturbation, 
and $\nu_{N,p}$ is the frequency with which a given neutral
atom collides with protons. With the above approximation to the collision 
integral, the solution to the Boltzmann equation reads
\begin{equation}
\delta\!f_N={(\bld{v}\cdot \bld{v}_N)\over{T/m_N}}
{n_N \exp({-m_Nv_N^2/2T})\over (2\pi T/m_N)^{3/2}}
{\nu_{N,p}\over \nu_{N,p}-i\omega+i\bld{k}
\cdot\bld{v}_N} \ .
\label{eq:fn}
\end{equation}

Next we verify that the perturbed neutral
number density vanishes,
\begin{eqnarray}
\int \delta\!f_N d^3\bld{v}_N&=&
{n_Nv\nu_{N,p}\over (2\pi)^{3/2} (T/m_N)^{5/2}} \nonumber \\ &\times&
\Big[
\int dv_{N\parallel}v_{N\parallel}\exp({-m_Nv_{N\parallel}^2/2T)}
\Big] \nonumber \\ &\times&
\Big[
\int d^2\bld{v}_{N\perp}
{\exp({-m_Nv_{N\perp}^2/2T)}
\over \nu_{N,p}-i\omega+i\bld{k}\cdot\bld{v}_{N\perp}}
\Big] \\
&=&0 \ ,
\end{eqnarray}
since the first square-bracketed integral vanishes by antisymmetry.
For the purpose of evaluating the above integrals, we use $\perp$
and $\parallel$ for the perpendicular and parallel projections
of $\bld{v}_N$
relative to $\bld{v}$, and not relative
to the magnetic field as in the rest of this paper; i.e.
$\bld{v}_{N\parallel}\equiv
(\bld{v}\cdot\bld{v}_N)\bld{v}/v^2$ and
$\bld{v}_{N\perp}\equiv\bld{v}_N-\bld{v}_{N\parallel}$.
We also use the incompressibility relation
$\bld{k}\cdot\bld{v}=0$, which implies that
$\bld{k}\cdot\bld{v}_{N\parallel}=0$.

With the neutral distribution function given by equation (\ref{eq:fn}),
the force per unit volume of the protons on the neutrals may
be calculated as
\begin{eqnarray}
\bld{F}
&=&m_N  \int  \bld{v}_N
(-i\omega \,\delta\!f_N+i(\bld{k}\cdot\bld{v}_N) \delta\!f_N)
 d^3\bld{v}_N \\
&=&
{m_Nn_N\bld{v}\nu_{N,p}\over (2\pi)^{3/2} (T/m_N)^{5/2}}\nonumber\\ &\times&
\Big[
\int dv_{N\parallel}v_{N\parallel}^2\exp({-m_Nv_{N\parallel}^2/2T)}
\Big] \nonumber \\ &\times&
\Big[
\int d^2\bld{v}_{N\perp}
{(-i\omega+i\bld{k}\cdot\bld{v}_{N\perp})\exp({-m_Nv_{N\perp}^2/2T)}
\over \nu_{N,p}-i\omega+i\bld{k}\cdot\bld{v}_{N\perp}}
\Big] \\
&=&
{m_Nn_N\bld{v}\nu_{N,p}\over (2\pi T/m_N)^{1/2}} \nonumber\\ &\times&
\int_{-\infty}^\infty ds {(-i\omega+iks)\exp({-m_Ns^2/2T)}
\over \nu_{N,p}-i\omega+iks} \\
&=&
{m_N n_N
\bld{v}\nu_{N,p} \over \pi^{1/2}\Pi_1}\nonumber\\&\times&
\int_{-\infty}^{\infty}
{ it \over
{1+it}  }
\exp \big[-(t/\Pi_1+\Pi_2/\Pi_1)^2\big]
dt  \ .
\end{eqnarray}
The second equality follows after replacing the overall multiplicative
$\bld{v}_N$
($=\bld{v}_{N\parallel}+\bld{v}_{N\perp}$)
by $\bld{v}_{N\parallel}$, because the
$\bld{v}_{N\perp}$ term vanishes by antisymmetry of
the integral over $v_{N\parallel}$.
In the second equality, the first bracketed integral
is equal to $(2\pi)^{1/2}(T/m_N)^{3/2}$.  The second
bracketed integral is a double integral;
the integral over the component
of $\bld{v}_{N\perp}$ which is perpendicular
to $\bld{k}$ is equal to $(2\pi T/m_N)^{1/2}$.
The remaining integral is over the component of
$\bld{v}_{N\perp}$ which is parallel to $\bld{k}$;
we label this component $s$ in the third equality.
Finally, the fourth equality follows from the change of
variables $t\equiv (ks-\omega)/\nu_{N,p}$, and
from the following definitions
\begin{eqnarray}
\Pi_1&\equiv& {k\over \nu_{N,p}} \Big({ {2T \over m_N}}\Big)^{1/2}
={v_A k \over \nu_{N,p}} {\beta}^{1/2}
\Big({m_p\over m_N}\Big)^{1/2}  \\
\Pi_2 &\equiv& {\omega\over\nu_{N,p}} =
{v_A k_\parallel\over \nu_{N,p}} \ ,
\end{eqnarray}
where $m_p$ is the proton mass.
The dimensionless parameter
$\Pi_1$ is the number of wavelengths a neutral crosses before
colliding, and the dimensionless parameter
$\Pi_2$ is the number of waveperiods a neutral
travels before colliding.
Since $k\gg k_\parallel$ and $\beta\gtrsim 1$, $\Pi_1\gg \Pi_2$
in the inertial range of the MHD cascade.

Performing the integral for the
real and imaginary parts of $\bld{F}$ to lowest order in
$\Pi_2/\Pi_1$ yields
\begin{equation}
\bld{F}=m_N n_N \bld{v}\nu_{N,p}
\Big(1-2i{\Pi_2\over \Pi_1^2}\Big)
g(\Pi_1) \ ,
\end{equation}
where
\begin{eqnarray}
g(\Pi_1)&\equiv& 1-
\sqrt{\pi}
\exp(\Pi_1^{-2}){\rm{erfc}}(\Pi_1^{-1})/\Pi_1 \label{eq:gdef}  \\
&=& \left\{
\begin{array}{lr} {1\over 2}\Pi_1^2 \ , \ \ & {\rm for}\ \Pi_1\ll 1 \\
1 \ , \ \ &{\rm for}\   \Pi_1\gg 1 \end{array} \right. \ .
\end{eqnarray}
Since $-\bld{F}$ is equal to the force of the neutrals on the protons,
we insert $-\bld{F}$ into the equation of momentum conservation
for the protons.  More precisely, we add $-i\bld{F}/(m_pn)$
to the right-hand side of equation (\ref{eq:pcons}) in the
Appendix.
We label the resulting change of frequency caused by
the presence of neutrals as $\Delta \omega$.
Assuming that $\Delta \omega\ll v_Ak_\parallel$, it 
follows that
\begin{equation}
{{\Delta \omega}\over v_Ak_\parallel}=-{ {m_N n_N} \over {m_p n} }
\Big({1\over \Pi_1^2}+{i\over {2\Pi_2}} \Big)g(\Pi_1) \ .
\label{eq:omeganeutral}
\end{equation}

\subsubsection{Discussion of the Frequency Change}

The physical interpretation of this frequency change is straightforward.
Recall that $k\gg k_\parallel$, so that the wavelength, $2\pi/k$, is
nearly identical to the wavelength transverse to the magnetic
field, $2\pi/k_\perp$.
It is convenient to define the neutral
mean free path, ${\rm L}_N$, as follows:
\begin{equation}
{\rm L}_N\equiv{\Pi_1\over k}={c_T\over \nu_{N,p}}
\Big( {2m_p\over m_N}\Big)^{1/2} \ ,
\label{eq:ln}
\end{equation}
where $c_T=(p/\rho)^{1/2}$ is the isothermal sound speed.

The imaginary part of $\Delta \omega$ is the negative of the damping
rate, and is given by
\begin{eqnarray}
\omega_i&=&-{1\over 2}{ {m_N n_N} \over {m_p n} }\nu_{N,p}g(k{\rm L}_N)
\label{eq:omegai} \\
&=&
-{1\over 2}{ {m_N n_N} \over {m_p n} }\nu_{N,p}
\times\left\{
\begin{array}{lr} {1\over 2}(k{\rm L}_N)^2 \ ,
\ \ & {\rm for}\ k{\rm L}_N\ll 1 \\
1 \ , \ \ &{\rm for}\   k{\rm L}_N\gg 1 \end{array} \right. \ .
\label{eq:omegai2}
\end{eqnarray}
The above expression may be explained as follows \citep{gs95}.
For wavelengths larger than the neutral mean free path,
$k{\rm L}_N < 1$, the neutrals are nearly locked to the protons.
They damp the MHD waves by transferring the protons' momentum across a
neutral mean free path. The damping rate increases with decreasing
wavelength provided $k{\rm L}_N \lesssim 1$. For wavelengths smaller than the neutral mean 
free path, the neutral atoms are effectively freely streaming. This leads 
to a damping rate which is independent of wavelength.

The real part of the frequency change, $\Delta \omega_r$, is
\begin{equation}
{{\Delta \omega_r}\over v_Ak_\parallel}=-{ {m_N n_N} \over {m_p n} }
\times\left\{
\begin{array}{lr} {1\over 2}  \ , \ \ & {\rm for}\ k{\rm L}_N\ll 1 \\
(k{\rm L}_N)^{-2} \ , \ \ &{\rm for}\   k{\rm L}_N
\gg 1 \end{array} \right. \ .
\label{eq:omegar}
\end{equation}
Its physical significance is apparent. For small wavelengths the neutral
atoms are effectively freely streaming,
the motion of the protons nearly decouples from the motion
of the neutrals, and there is negligible real frequency change
associated with the presence of the neutrals.
For large wavelengths the neutrals are locked to the protons, so
the mass density of the fluid which participates in the waves is larger
than that of the protons.
Since the Alfv\'en speed, and hence the wave frequency, is
inversely proportional to the square root of the mass density,
i.e. $\omega\propto v_A\propto \rho^{-1/2}$, this results in a decrease of the 
frequency by a
fractional amount which is equal
to one-half of the mass ratio of neutrals to protons.

\subsection{Effect of Neutrals on the Turbulent Cascade}

\subsubsection{Effect of $\omega_i$ on the Cascade}
The turbulent cascade is quenched if the damping rate exceeds
the eddy cascade rate or, equivalently, if
$\vert\omega_i\vert/v_Ak_\parallel>1$.
From equation (\ref{eq:omegai2}) and $k_\parallel\propto k^{2/3}$,
it is seen that
for small wavenumbers $\vert\omega_i\vert/v_Ak_\parallel\propto k^{4/3}$, 
and for large wavenumbers $\vert\omega_i\vert/v_Ak_\parallel\propto k^{-2/3}$.
Therefore $\vert\omega_i\vert/v_Ak_\parallel$
is a maximum for transverse wavelengths, $k^{-1}$,
comparable to the neutral
mean free path, and decreases for both
larger and smaller wavelengths.  
The requirement that the cascade survive damping by neutrals
is then that $\vert\omega_i\vert/v_Ak_\parallel<1$ for $k^{-1}$ comparable to the 
neutral mean free path,
or equivalently, $n_N\nu_{N,p}/n\lesssim v_Ak_\parallel$ at this
scale (eq. [\ref{eq:omegai2}]).  Since $n_N\nu_{N,p}/n$ is
the rate at which a given proton collides with neutrals,
the cascade survives to small scales if the wave frequency,
and hence the cascade rate,
at the scale of the neutral mean free path is faster than
the rate at which a proton collides with neutrals.

To obtain the total decrement in the amplitude of
both the Alfv\'en mode and the slow mode through the damping
scale, we solve the kinetic equation for the cascade.\footnote{
The calculation of the amplitude decrement in
this section is similar to that given in \S
\ref{sec:slowdamp} for slow mode damping at the cooling scale.  The principal 
difference, aside from the fact
that the imaginary part of the frequency is different,
is that here we must account for the decrease in the cascade rate, 
$t_{\lambda_\perp}$,
caused by damping.  This is not necessary when treating slow mode damping
at the cooling scale because there the Alfv\'en modes which control the
cascade rate are undamped.}
As in \S \ref{sec:slowdamp}, the kinetic equation is obtained by
balancing the $k$-space energy flux with the loss-rate of
$k$-space energy density due to damping:
\begin{equation}
{d\over {dk}} {v_{\lambda_\perp}^2\over t_{\lambda_\perp}}
=2\omega_i {v_{\lambda_\perp}^2 \over k} \ ,
\end{equation}
where $\lambda_\perp^{-1}\equiv k_\perp\simeq k$ is the 
transverse wavenumber, and
$t_{\lambda_\perp}$ is the cascade time.
The kinetic equation can be rewritten as follows:
\begin{equation}
{d\over d\ln k}
\ln {v_{\lambda_\perp}^2\over t_{\lambda_\perp}}
=2M_t{\omega_i\over v_A k_\parallel}  \ ,
\label{eq:kinetic}
\end{equation}
where  $M_t$ is a Kolmogorov constant (see \S \ref{sec:kolmogorov}).
Before integrating this equation, we re-express the
damping frequency in terms of the relevant lengthscales,
$\lout$ and ${\rm L}_N$:
\begin{eqnarray}
{\omega_i\over v_A k_\parallel}
&=&-{1\over 2}{m_Nn_N\over m_pn}
\nu_{N,p}
{g(k{\rm L}_N)\over v_A k_\parallel} \\
&=&
-{1\over 2M_\parallel}{m_Nn_N\over m_pn}
\nu_{N,p}
{g(k{\rm L}_N)\over v_{\lambda_\perp} k} \\
&=&
-{1\over \sqrt{2}M_\parallel}\Big({m_N\over m_p}\Big)^{1/2}{n_N\over n}
\Big({\lout\over{\rm L}_N}\Big)^{1/3}
{g(k{\rm L}_N)\over (k{\rm L}_N)^{2/3}}  \ .
\label{eq:om}
\end{eqnarray}
The first equality above follows from equation (\ref{eq:omegai}).
The second equality follows from the Kolmogorov constant
$M_\parallel=v_Ak_\parallel/v_{\lambda_\perp} k$
(see \S \ref{sec:kolmogorov}).
The third equality follows from the velocity spectrum,
$v_{\lambda_\perp}=c_T(k\lout)^{-1/3}$ (eq. [\ref{eq:vspec}]), and from
the definition of the neutral mean free path (eq. [\ref{eq:ln}]).
For convenience, we define $\lout$ in this section such that 
$v_{\lambda_\perp}=c_T$ when $k^{-1}=\lout$.

Integrating the kinetic equation
(eq. [\ref{eq:kinetic}]), and using
$t_{\lambda_\perp}\sim \lambda_\perp/v_{\lambda_\perp}$,
yields the net decrement due to damping by collisions with neutrals:
\begin{eqnarray}
{[v_{\lambda_\perp}/\lambda_\perp^{1/3}]\vert_{\lambda_\perp\ll {\rm L}_N}
\over
[v_{\lambda_\perp}/\lambda_\perp^{1/3}]\vert_{\lambda_\perp\gg {\rm L}_N}}
&=&
\exp\Big[
{2M_t\over 3}\int_0^\infty
{\omega_i\over v_A k_\parallel}{dk\over k}
\Big] \nonumber\\
&=&\exp\Big[
-0.2\Big({m_N\over m_p}\Big)^{1/2}{n_N\over n}
\Big({\lout\over{\rm L}_N}\Big)^{1/3}
\Big] \ ,
\label{eq:neutraldamp}
\end{eqnarray}
after substituting from equation (\ref{eq:om}). The numerical prefactor
in the second equality, $0.2$, follows after inserting the values
of the Kolmogorov constants (eq. [\ref{eq:kolmogorov}]) and
the value of the integral
\begin{equation}
\int_0^\infty g(x)x^{-5/3}dx\simeq 1.1 \ ,
\end{equation}
which was integrated numerically; see equation (\ref{eq:gdef}) for the
definition of $g(x)$.

For the cascade to continue to small scales, the right-hand side
of equation (\ref{eq:neutraldamp}) cannot be very small.
This can be viewed as an upper limit on the neutral fraction:
$n_N/n\lesssim ({\rm L}_N/\lout)^{1/3}$.
Recall that the condition for the cascade to reach small scales
is that the cascade rate at ${\rm L}_N$ be faster than
the rate at which a proton collides with neutrals.  If
the neutral fraction is too large, then so is the proton collision
rate, and the cascade is quenched.  Moreover, for a fixed
${\rm L}_N$, a large value for $\lout$ implies that the cascade
time at ${\rm L}_N$ is large, and hence that the cascade is
more susceptible to damping by neutrals.

We postpone consideration of the damping due to hydrogen and helium atoms until
after we evaluate the effect of $\Delta\omega_r$ on the cascade.

\subsubsection{Effect of $\Delta\omega_r$ on the Cascade}

We can picture the cascade as proceeding from large scales to small scales.
As it crosses the scale of the neutral mean free path (eq. [\ref{eq:omegar}]), 
the effective Alfv\'en speed, and hence the real part of wave frequency,  
increases by the fraction $m_Nn_N/2m_pn$. Consequently, the cascade time 
decreases by the same amount.  To the extent that the flux of energy in 
Alfv\'en waves from large scales to small scales is constant in the turbulent 
cascade, $v_{\lambda_\perp}^2/t_{\lambda_\perp}=$ constant, 
where $t_{\lambda_\perp}$ is the cascade
time. Therefore, a fractional decrease of $m_Nn_N/2m_pn$ in the cascade time 
causes a decrement in the small-scale Alfv\'enic velocity perturbation by the 
fractional amount $m_Nn_N/4m_pn$.  This decrement is in addition to that due to 
damping. Moreover, it applies to the slow mode and the entropy mode as well
as to the Alfv\'en mode. Because $n_N/n$ must be small for the cascade to pass
through the scale of neutral damping, this decrement is also small, and we
ignore it from here on.

\subsection{Neutral Hydrogen Atoms}
When the neutrals are hydrogen atoms
we set
$n_N=n_{{\rm H}}$, $\nu_{N,p}=\nu_{{{\rm H}},p}$,
${\rm L}_N={\rm L}_{\rm H}$, and $m_N=m_p$.
Collisions between hydrogen atoms and protons are due
to resonant charge exchange.
From equation (\ref{eq:ln}), with
the value of $\nu_{{\rm H},p}$ taken from \citet{banks66},
\begin{equation}
{\rm L}_{\rm H}=5\times 10^{13}
\Big({{\rm cm}^{-3}\over n}\Big)
\ \ {\rm cm} \
\end{equation}
at $8\times 10^3\,$K; the temperature dependence of ${\rm L}_{\rm H}$
is very weak.

For the cascade to survive damping by neutral hydrogen atoms,
the right-hand side of equation
(\ref{eq:neutraldamp}) cannot be very small.  This sets an upper limit
on the neutral fraction of
\begin{equation}
{n_{{\rm H}}\over n}
\lesssim  5\Big({{\rm L}_{{\rm H}}\over \lout}\Big)^{1/3}
\sim \Big({6\times 10^{15}
{\rm cm}\over (n/{\rm cm}^{-3})\lout}\Big)^{1/3} \ .
\label{eq:neutralfrac}
\end{equation}

The real part of $\omega/v_A k_\parallel$ is larger on scales below ${\rm 
L}_{{\rm H}}$ than it is on scales above ${\rm L}_{{\rm H}}$ by the fractional 
value $n_{{\rm H}}/2n$.

\subsection{Neutral Helium Atoms}

Although helium has a lower abundance than hydrogen,
it has a higher ionization potential.  Therefore, in
some regions helium might comprise the majority of
neutrals.
When the neutrals are helium atoms we set
$n_N=n_{{\rm He}}$, $\nu_{N,p}=\nu_{{{\rm He}},p}$,
${\rm L}_N={\rm L}_{\rm He}$, and $m_N=4m_p$.
From equation (\ref{eq:ln}), with the value of
$\nu_{{\rm He},p}$ taken from \citet{banks66},
\begin{equation}
{\rm L}_{{\rm He}}=
1.5\times 10^{15}
\Big({{\rm cm}^{-3}\over n}\Big)
\ \ {\rm cm} \
\end{equation}
at $8\times 10^3\,$K, and ${\rm L}_{{\rm He}}\propto T^{1/2}$.

To place an upper bound on damping by helium atoms,
we assume that most of the helium is neutral, i.e.
$n_{{\rm He}}\approx 0.1 n$. Then the decrement is given
by the right-hand side of equation (\ref{eq:neutraldamp}):
\begin{equation}
{[v_{\lambda_\perp}/\lambda_\perp^{1/3}]\vert_{\lambda_\perp\ll 
{\rm L}_{{\rm He}}}
\over
[v_{\lambda_\perp}/\lambda_\perp^{1/3}]\vert_{\lambda_\perp\gg 
{\rm L}_{{\rm He}}}}
=
\exp\Big[
-\Big(
{(n/{\rm cm}^{-3})\lout\over 2\times 10^{19} {\rm cm}}
\Big)^{1/3}
\Big] \ .
\end{equation}
Only if $\lout\gtrsim 2\times 10^{19} (\cm^{-3}/n) \cm$ could
the cascade be terminated at the scale of the helium
mean free path.  If there are not many neutral helium atoms,
or if the outer scale is not sufficiently large, then
damping at ${\rm L}_{{\rm He}}$ may be neglected, and the cascade
extends at least to the scale of the hydrogen mean free path.

Collisions of neutral helium atoms with singly ionized helium ions
might also be significant.
Although there are fewer helium atoms than
protons, He$^0$-He$^+$ collisions have a larger cross-section than
He$^0$-proton collisions because they are due to resonant charge
exchange.
Nonetheless, given the cosmic abundance of helium, 
the mean free path of neutral helium due to He$^0$-He$^+$ collisions cannot
be significantly smaller than that due to He$^0$-proton collisions 
regardless of 
the ionization fraction of helium. 

The real part of $\omega/v_Ak_\parallel$ increases by
$2n_{{\rm He}}/n$
below the scale ${\rm L}_{{\rm He}}$,
or 0.2 if most of the helium is neutral.

\subsection{If Neutrals Damp the Cascade}

In subsequent sections, we consider
regions in the interstellar medium where there
are too few neutrals to damp the
cascade.  However, there are almost certainly
many regions where
the neutrals do damp the cascade.
We discuss these regions here.

Suppose both Alfv\'en waves and slow waves are damped by neutrals. What happens 
to the entropy waves?  If undamped, they would be mixed down to smaller scales 
by Alfv\'en waves at the neutral damping scale. The resulting density spectrum
would be $n_{\lambda_\perp}=$ constant 
(see eq. [\ref{eq:cspectrum3}]). Because it is 
flatter than $n_{\lambda_\perp}\propto \lambda_\perp^{1/3}$, 
regions in which damping by 
neutrals truncated the Alfv\'en cascade might be important contributors to small 
scale density fluctuations. However, the fact that they would contribute a 
density spectrum different from that which is observed suggests that the entropy 
wave cascade is not more resistant than the Alfv\'en wave cascade to damping by 
neutrals. Indeed that is the case. 
Recall that the condition for the truncation of Alfv\'en and
slow wave cascades is that each proton collide with at least one
neutral atom during one wave period at the neutral damping scale.  
Under this condition, the neutrals would damp the entropy waves by conducting 
heat across them.

Although regions in which the cascade is damped by neutrals do not contribute
small scale density fluctuations, they may still be significant. Observational 
evidence indicates that there is more power in density fluctuations on large 
scales, $10^{13}-10^{14}\,$cm, than would be predicted by extrapolating from 
small scales, $10^8-10^{10}\,$cm, with the Kolmogorov scaling. See, for example, 
\citet{lr00} for a review of the observations. Perhaps this excess arises in 
regions where the cascade is damped by neutrals.

We complete this section by briefly considering and then rejecting the 
possibility that a turbulent cascade truncated at the neutral damping scale 
might be regenerated on a much smaller scale due to stirring by eddies at the 
damping scale. Although the ratio of the damping rate to the wave frequency 
decreases below the scale of the neutral mean free path, the absolute damping 
rate approaches a constant value. Provided the cascade is truncated by neutral 
damping, this rate is larger than the stirring rate and the cascade 
cannot be regenerated.

\section{THE COLLISIONLESS SCALE OF THE IONS}
\label{sec:collion}

If the cascade survives the neutral collisionless scales,
then, proceeding to smaller scales, the next scale of
importance is the ion collisionless scale.  This
scale is set by the mean free path of protons to collide
with other protons:
\begin{equation}
\lmfp=6\times 10^{11} \Big({{\rm cm}^{-3}\over n}
\Big) \ \ {\rm cm}
\end{equation}
at a temperature of 8,000K \citep{brag65}.
Since the electron and proton gyroradii
(see eq. [\ref{eq:protongyro}] for protons) are very
small compared to the proton mean free path, the electrons
and protons are tied to magnetic fieldlines.  Therefore,
when considering the collisionless effects of electrons and
protons, the relevant lengthscales are those parallel
to the magnetic field.
Since turbulent eddies are highly elongated along the
magnetic field, their transverse
lengthscales are much smaller 
than their parallel ones (eq. [\ref{eq:spectrum2}]).
It is the transverse lengthscale which is relevant when considering
observations of the density spectrum, because each line of sight
averages over regions with different magnetic field orientations. 
In this section, we show that
both the slow mode and the entropy
mode are cut off at the lengthscale where the parallel eddy size
is comparable to the proton mean free path.
The lengthscale where the density spectrum is {\it observed} to cut off,
i.e. the
transverse lengthscale, is therefore significantly smaller than
the mean free path.

Before discussing the damping of the slow mode and the entropy mode,
we consider two larger lengthscales, which are set by
the effects of the electrons.
Although the behaviour of the cascade at these lengthscales is
interesting, it is shown to be unimportant for
our purposes.

Throughout this section, we neglect numerical factors
of order unity, such as the factors which are associated with
kinetic corrections to the fluid equations (given in
Braginskii 1965)
and the Kolmogorov constants.  However, we retain the dependences
on $\beta$, which is assumed to satisfy $\beta>1$, and on
the ratio of the proton mass to the electron mass: $m_p/m_e=1840$.

\subsection{The Electron Diffusion Scale}

The electrons have the same mean free path as the protons, but they
are faster than the protons by the square root of the mass ratio:
\begin{equation}
c_{s,e}\approx\Big({m_p\over m_e}\Big)^{1/2}c_s \ ,
\end{equation}
where $c_{s,e}$ is the electron thermal speed and $c_s$ is the
sound speed, which is comparable to the proton thermal speed.
Because of charge neutrality, the electrons have
the same density, both perturbed and unperturbed, as the protons.
Viscous damping caused by the electrons may be neglected:
since the dynamic viscosity of the electrons is smaller than that of
the protons by the square root of their mass ratio, 
electron viscous damping is always subdominant.

As we now show, electrons are important for conducting heat
on parallel lengthscales slightly larger than the proton mean free path.
Electrons diffuse parallel to the magnetic
field across a distance $\lpar$ in the time
$(\lmfp/c_{s,e})(\lpar/\lmfp)^2$.
This is equal to
the cascade time of an eddy
with parallel lengthscale $\lpar$,
i.e. it is equal to $\lpar/v_A$, at the
``electron diffusion scale'', given by
$\lpar\sim\led$, where
\begin{equation}
\led\approx
{c_{s,e}\over v_A}\lmfp \approx
\beta^{1/2}\Big({m_p\over m_e}\Big)^{1/2}\lmfp \ .
\label{eq:led}
\end{equation}
In eddies with parallel lengths smaller than this, the electrons
diffuse across the parallel lengths of the eddies, thereby conducting heat,
and the electrons are isothermal;
in eddies with parallel lengths
larger than this, conduction is unimportant.

We must also consider the effect of electron conduction on the protons.
Electrons and proton temperatures approach a common value on the timescale
that there are
$m_p/m_e$ collisions per particle.  Since electrons are faster, the
collision time is set by the electron speed, and the time for
electron and proton temperatures to equilibrate is
\begin{equation}
\tau_{\rm eq}\approx {m_p\over m_e}{\lmfp\over c_{s,e}}
\approx {1\over v_A}
{1\over{\beta}^{1/2}}\Big({m_p\over m_e}\Big)^{1/2}\lmfp\ .
\label{eq:equil}
\end{equation}
This time is smaller, by a factor $\beta$, than the cascade time of
eddies with parallel size $\led$.
Thus in eddies of this parallel size, and in those which are
slightly smaller, protons are at the same temperature
as the electrons.  And, since the electrons are isothermal, so
are the protons.

As a result of the above considerations, when the parallel cascade
crosses the electron diffusion scale, the cascade becomes isothermal.
This is similar to the crossing of the cooling scale, discussed in
\S \ref{sec:coolingsmallbeta}, though backwards, and similar
considerations apply.  In particular, the Alfv\'en mode is incompressible,
and hence unaffected.  The slow mode is
nearly incompressible, and is only slightly affected: it
suffers some damping to first order in $1/\beta$.  The entropy
mode, however, does not exist under isothermal conditions;
as the electron diffusion scale is crossed, entropy waves are
converted into slow waves, and the density fluctuations which
had been associated with entropy waves now become associated
with slow waves.

\subsection{The Electron-Proton Equilibration Scale}

Continuing to slightly smaller scales, the isothermal cascade
reaches the ``equilibration scale'', where the cascade time,
$\lpar/v_A$,
is comparable to the electron-proton equilibration time
(eq. [\ref{eq:equil}]). At this scale, $\lpar\sim\lequ$, where 
\begin{equation}
\lequ\approx
{1\over{\beta}^{1/2}}\Big({m_p\over m_e}\Big)^{1/2}\lmfp\ .
\label{eq:lequ}
\end{equation}
On larger scales, the protons
are thermally coupled to the electrons, and hence they are isothermal
on slightly larger scales; on
smaller scales, the protons are thermally independent
of the electrons, and hence adiabatic.  The transition through
the equilibration scale is nearly identical to the transition through the
cooling scale (\S \ref{sec:coolingsmallbeta}): the Alfv\'en waves
are unaffected, and they mix larger-scale
isothermal slow waves into smaller-scale adiabatic slow
waves and into smaller-scale entropy waves, although with some
damping.
Therefore, below the equilibration scale, the entropy mode reappears.

The result of the calculations in both this subsection
and the previous subsection is that, in
crossing the electron diffusion scale and the equilibration
scale, density fluctuations which were associated with the entropy
mode on large scales are transferred from the entropy mode
to the slow mode and then back to the entropy mode. Density fluctuations
which were associated with the slow mode on large scales are
unaffected.\footnote{However, if the
density fluctuations in the large-scale slow mode were
less than the density fluctuations
in the large-scale entropy mode, then, on small scales, the slow mode
would be boosted so that its density fluctuations would be comparable
to those of the entropy mode: the slow mode can steal approximately half
of the entropy mode.}
Therefore, for our purposes, these two lengthscales have little net effect
on the density spectrum.
Although there is some damping, the amount of damping is
comparable to the amount at the cooling scale, and hence is not
very significant.

\subsection{The Proton Diffusion Scale: Death of the Slow Mode and
Entropy Mode}

Entropy waves and slow waves with parallel wavelengths smaller than
the proton mean free path both
damp on the timescale that protons stream across their parallel wavelengths
\citep{barnes66}.  However, because we consider $\beta>1$, 
there
is a scale slightly larger than the proton mean free path at which
these waves damp in the turbulent cascade.  This is the proton diffusion scale,
which is the scale at which 
protons can diffuse across an eddy within a cascade time.
Below the proton diffusion scale, proton viscosity kills the slow waves
and proton heat conduction kills the entropy waves.
Alfv\'en waves are unaffected by either of these effects.  
The density spectrum therefore cuts off below the proton diffusion scale.

To evaluate the proton diffusion scale, we equate
the time for protons to diffuse
across an eddy of parallel lengthscale $\lpar$, i.e. 
$(\lmfp/c_s)(\lpar/\lmfp)^2$,
with the cascade time, $\lpar/v_A$.  This gives
$\lpar\sim\lpd$, where
\begin{equation}
\lpd\approx
{c_s\over v_A}\lmfp \approx
\beta^{1/2}\lmfp \ .
\end{equation}
However, from an observational point of view, it is the transverse
size of a damped eddy, $\lambda_\perp$, which would be observed.
This is related to the parallel size through equation (\ref{eq:spectrum2}).
Alternatively, we use the Kolmogorov constant
$M_\parallel\equiv v_A\lambda_\perp/v_{\lambda_\perp}\lpar$
(see \S \ref{sec:kolmogorov}).\footnote{
Although we drop other factors of
order unity, we keep the dependence on this one because of the
importance of the proton diffusion scale.}
Then, the transverse scale of a damped eddy, i.e.
an eddy with parallel size $\lpar\sim\lpd$, is
\begin{eqnarray}
\lambda_\perp&\approx&M_\parallel
{v_{\lambda_\perp}\over v_A}
\lpd \nonumber \\
&\approx&M_\parallel\beta\lmfp
\Big({\lambda_\perp\over \lout}\Big)^{1/3} \ ,
\end{eqnarray}
which we solve for the cut-off lengthscale: 
$\lambda_\perp\sim \lpdperp$, where
\begin{eqnarray}
\lpdperp&\approx& M_\parallel^{3/2}
\beta^{3/2}\lmfp
\Big({\lmfp\over \lout}\Big)^{1/2} \\
&\approx& 2\times 10^9 
\Big({{\rm pc}\over \lout}\Big)^{1/2}
\Big({\beta\over n/{\rm cm}^{-3}}\Big)^{3/2} \ \ \cm 
\ .
\end{eqnarray}
Below this scale, the density spectrum is cut off.
For plausible values of
$\beta$, $\lout$, and $n$, this lengthscale is significantly
smaller than $\lmfp$.  It is also larger than the proton
gyroradius (eq. [\ref{eq:protongyro}], below).

\citet{ars95} summarize the observations of the density spectrum
cut-off.  There is considerable evidence that the cut-off scale is smaller
than about $10^{10}$ cm along many directions.
There is weaker evidence, from refractive
scintillation, that along some lines of sight the cut-off scale
is larger than around $10^9$ cm.  Our theory might have implications
for these observations.

We conclude this section with two remarks.  First, we re-emphasize
the importance of the fact that the cascade is anisotropic.
It is this fact which allows the density spectrum to reach lengthscales
which are significantly smaller than the proton mean free path.
Second, we note an important consequence of the parallel
cascades of the slow mode and of the passive scalar;
these parallel cascades were explained
in great detail when considering incompressible MHD turbulence
(\S \ref{sec:incomp}).
Had there been no parallel cascade, 
then neither the slow
mode nor the entropy mode would damp at the proton
diffusion scale.  The wavelength along the magnetic field
would be effectively infinite, and so the ions could not diffuse
across wavelengths.  Rather, the slow mode and the entropy mode would
be mixed down to the proton gyro scale, where the Alfv\'enic cascade
is cut off.

\subsection{Density Spectrum Below the Proton Diffusion Scale}

On scales smaller than $\lpdperp$, density fluctuations are wiped
out.  Since on these scales protons diffuse across 
the lengths of many eddies 
before the eddies cascade, the density within neighbouring eddies
is homogenized.  
Homogenization occurs not only 
in the direction parallel to 
the local mean magnetic field---as
might have been expected since protons are tied to fieldlines---but
also in the transverse direction.  
This is because eddies that are adjacent in the 
parallel direction incorporate
substantially different fieldlines, so proton diffusion
also wipes out density differences amongst eddies with
transverse separations.
The result is that the density spectrum on scales smaller 
than $\lpdperp$ is determined by density fluctuations at 
$\lpdperp$; equivalently,
\begin{equation}
n_{\lambda_\perp}\propto \lambda_\perp \ \ , \ \
{\rm for}\ \lambda_\perp<\lpdperp  \ .
\end{equation}

\section{THE END: THE PROTON GYRO SCALE}

The Alfv\'en mode is undamped
in a collisionless medium \citep{barnes79}.
Thus, the Alfv\'en wave  cascade survives below the proton
diffusion scale, without the accompaniment of the slow waves and
entropy waves.
The Alfv\'enic cascade is damped at the scale of the proton gyroradius:
\begin{equation}
{\rm L}_{\rm p,gyr}=\sqrt{2}c_T{m_pc\over eB}=
2.5\times 10^7 \Big({\beta\over n/{\rm cm}^{-3}}\Big)^{1/2} \ \ \cm \ .
\label{eq:protongyro}
\end{equation}
At this scale, Alfv\'en
waves are converted into whistlers.  The whistlers cascade
to smaller scales, where they 
are damped by
the collisionless effects of the electrons 
(Quataert 1998, viz. the curve in his Fig. 1b that corresponds
to equal electron and proton temperatures).

\section{SUMMARY}

The primary goal of this paper has been to calculate the small-scale
density spectrum in turbulent interstellar plasmas.
Our theory of compressible MHD turbulence is based upon the
incompressible theory of \citet{gs95}, for which there is growing
support from numerical simulations.
We hypothesize that the compressible theory
is similar to the incompressible theory of MHD turbulence, but with two main
modifications:
a compressible slow mode and an entropy mode which is passively
advected.
While we believe that this hypothesis is plausible, it can, and should,
be tested with numerical simulations of compressible MHD.

Because of the multitude of special lengthscales encountered
when discussing the turbulent cascade in the interstellar medium,
a recapitulation might be useful.  In the following, we list
the most important lengthscales, and summarize their significance to the
turbulent cascade.

\noindent {\em The Outer Scale} ($\lout$): This is the lengthscale at which
the turbulent motions are stirred. 
On slightly smaller scales, there is 
isothermal hydrodynamic turbulence
for most plausible astrophysical sources.

\noindent {\em The MHD Scale} ($\lmhd$): At this lengthscale there
is a transition from hydrodynamic
to MHD turbulence.  Larger-scale hydrodynamic motions couple to
smaller-scale Alfv\'en waves and slow waves.

\noindent {\em The Cooling Scale} ($\lcool$):
At this lengthscale there is a transition from isothermal to
adiabatic turbulence.  
In high-$\beta$ turbulence, where small scale density
fluctuations are due to the entropy mode, entropy---and hence
density---fluctuations are suppressed by cooling.  
In low-$\beta$
turbulence, i.e. $1\lesssim\beta<\lout/c_st_{\rm cool}$,
density fluctuations due to the slow mode
are important, and cooling has a negligible effect on the small-scale
density spectrum.

\noindent {\em The Collisionless Scale of the
Neutrals} (${\rm L}_{\rm H}$; ${\rm L}_{\rm He}$):
Neutrals decouple from ions across these scales.
If the neutral
fraction is not sufficiently small, then all three modes---Alfv\'en,
slow, and entropy---are damped.

\noindent {\em The Collisionless Scale of the Ions}
($\lpar=\led$; $\lpar=\lequ$; $\lpar=\lpd
\Leftrightarrow \lambda_\perp=\lpdperp$):
Across these lengthscales there is a gradual transition from fluid behaviour
to collisionless plasma behaviour.  The first two of these lengthscales,
the electron diffusion scale and the equilibration scale,
are set by the electrons. These lengthscales
have only a small net effect on the cascade.  The proton diffusion
scale, however, is critically important for density fluctuations:
both the slow mode and the entropy mode are cut off at this scale,
and hence there are no density fluctuations below $\lpdperp$.

\noindent {\em The Proton Gyro Scale} (${\rm L}_{\rm p,gyr}$):
Alfv\'en waves are
cut off at this scale.  However, this scale has little importance
for the density spectrum, because there are no density fluctuations below
the proton diffusion scale.

\section{COMPARISON WITH HIGDON'S WORK}
\label{sec:higdon}

In his 1984 paper, Higdon attributes small-scale density fluctuations
to the passive mixing of the entropy mode.  In his 1986 paper, Higdon
notes that if the entropy mode varies along the magnetic field, it
is damped in a collisionless medium.
He then attributes small-scale density fluctuations to the passive
mixing of tangential pressure balances.  Tangential pressure
balances are structures which are parallel to the mean magnetic field.
They are composed of both entropy waves and slow waves that have purely
transverse wave vectors.

Considering that Higdon's papers preceded even a theory of incompressible 
MHD turbulence, they are a remarkable accomplishment.  
However, Higdon does not account for the parallel cascade. 
Entropy and slow waves with purely
transverse wave vectors contain negligible power.
They are cascaded along the magnetic field by Alfv\'en waves.
Consequently, they are damped when protons can diffuse across
eddies in a cascade time, i.e. below the proton diffusion scale.

\section{COMPRESSIBLE TURBULENCE WHEN $\beta<1$}
\label{sec:betalt1}

In this section only, we consider compressible turbulence 
in plasmas that have $\beta<1$, e.g. in isothermal shocks
and in the solar wind.

Since Alfv\'en waves are unaffected by the value of $\beta$, and
since nearly transverse slow waves are only slightly affected
(see Appendix), the dynamics of the cascade is nearly 
independent of $\beta$.  
While the dispersion relation of the slow mode is changed from 
$\omega=v_Ak_z$ to $\omega=c_sk_z$, the slow waves are still
passively mixed by the Alfv\'en waves.
Therefore, the Alfv\'en, slow, and entropy spectra 
in $\beta<1$ MHD turbulence are the same as when $\beta>1$.

However, the damping of the slow and entropy waves
is significantly changed.
When $\beta>1$, the proton thermal speed is faster than
the Alfv\'en speed. Therefore protons can stream across small
eddies before they cascade.
Conversely, when $\beta<1$, the time for protons to cross an eddy
is always shorter than the cascade time.  Therefore slow and
entropy waves
cannot be damped by protons which cross eddies, and the density
spectrum extends to smaller scales.

We have ignored the effects of the electrons.
However, in the following we show that the electrons' behaviour
may be ignored for our purposes.
Our discussion closely parallels that 
when $\beta>1$ in \S \ref{sec:collion}, and uses similar notation.
When $\beta<1$, the equilibration scale
(eq. [\ref{eq:lequ}]), is larger than the 
electron diffusion scale (eq. [\ref{eq:led}]).
Thus the largest scale at which kinetic effects are significant
is the scale at which the cascade time is comparable to
the time for electrons and protons to equilibrate their temperatures.
Below this scale, electrons and protons are thermally decoupled.
Nonetheless, this thermal decoupling has no effect on the Alfv\'en,
slow, or entropy waves on scales larger than 
the electron diffusion scale.  At the electron diffusion scale,
the cascade
time is comparable to the electron diffusion time.  Below this
scale, the electrons are isothermal.  However, since the protons
are thermally decoupled from the electrons, the slow waves and entropy waves
are cascaded to smaller scales.\footnote{There is a change in the
density spectrum at this scale which is of order unity; it is due to
the change in the electrons' equation of state.}

Continuing to smaller scales, the next scale of importance is
that at which the parallel size of an eddy is comparable to
the proton mean free path.  Below this scale, the cascade is
collisionless.  The Alfv\'en waves are undamped
by collisionless effects. The entropy waves are undamped
since protons cannot cross eddies within a cascade time.
The slow waves are also undamped within a cascade time: although they 
damp within a waveperiod by Barnes damping,
their waveperiod is longer than the cascade time.
Consequently, the Kolmogorov density spectrum extends down to the
proton gyro scale.

\section{FUTURE WORK}

In a future paper, we will examine in detail the 
density spectrum in the solar wind.
In another paper (paper II), we will relate the theory of compressible plasma 
turbulence developed here to observations---primarily those 
of diffractive scintillation.  We will demonstrate that 
the observed amplitude of small-scale density fluctuations is surprisingly 
large, especially along certain lines of sight, such as the one toward
the Galactic Center. Then we will attempt to determine
which astrophysical sources contribute the bulk of the density fluctuations and 
why they do so.

\acknowledgments
We thank Jason Maron for analyzing some of his simulations for us,
for showing us how to use his numerical code, and
for informative discussions.
Research reported in this paper was supported by NSF grant 94-14232.

\appendix
\section{WAVES IN COMPRESSIBLE MHD}

In this Appendix we derive the properties of the Alfv\'en mode,
the slow mode, and the entropy mode.  The fast mode is not relevant  to
interstellar scintillation
for reasons discussed in the body of the paper.
The Fourier-transformed, linearized equations of ideal MHD, with 
$\partial/\partial t\rightarrow -i\omega$ and $\bld{\nabla}\rightarrow 
i\bld{k}$, read
\begin{eqnarray}
\omega n'&=&(\bld{k}\cdot\bld{v})n\, ,
\, \qquad\mbox{(mass conservation)} \\
\omega \bld{v}&=&\bld{k}
\Big[\tilde{c}^2{n'\over n}
+b_z v_A\Big]-\bld{b}v_A k_z
\, , \qquad\mbox{(momentum conservation)} \label{eq:pcons} \\
\omega\bld{b}&=&\bld{\hat{z}}v_A(\bld{k}\cdot\bld{v})
-\bld{v}v_A k_z \ ,
\, \qquad\mbox{(Faraday's law)}
\end{eqnarray}
where $n$ is number density, $\rho$ is mass density, $B$ is the background 
magnetic field intensity, and $v_A\equiv B/\sqrt{4\pi\rho}$ is the Alfv\'en 
speed. The $z$-axis is chosen to lie parallel to the
background magnetic field. Some other variables are the perturbed number 
density, $n'$, the fluid velocity, $\bld{v}$, and the perturbed magnetic 
intensity, $\bld{B}'$, or in velocity units, $\bld{b}\equiv 
\bld{B}'/\sqrt{4\pi\rho}$. The sound speed, $\tilde{c}$, is defined less 
conventionally here to be the square root of the ratio
of perturbed pressure to perturbed mass density,
$\tilde{c}\equiv ({p'/\rho'})^{1/2}$.
For an isothermal gas, $\tilde{c}=c_T$,
where $c_T$ is the isothermal sound speed;
$c_T^2=p/\rho$. 
For an adiabatic ideal gas,
$\tilde{c}=c_s$, where $c_s$ is the adiabatic sound speed; $c_s^2=\gamma 
p/\rho=5p/3\rho$ for a monatomic plasma. 
More generally, however, the
equation of energy conservation must be specified before
$\tilde{c}$ can be determined.
While in general $\tilde{c}$ depends
on $\omega$, we ignore the $\omega$-dependence in this Appendix,
and treat $\tilde{c}$ as a constant. This treatment is valid
for the purposes of this paper.

We can express the equations of motion in terms of $\bld{v}$ as follows:
\begin{equation}
(\omega^2-k_z^2v_A^2)\bld{v}
=[(\bld{k}\cdot\bld{v})(\tilde{c}^2+v_A^2)-k_zv_zv_A^2]\bld{k}
-v_A^2k_z(\bld{k}\cdot\bld{v})\bld{\hat{z}} \ .
\label{eq:eomexpand}
\end{equation}

\subsection{Alfv\'en Mode}

The Alfv\'en mode is incompressible; $\bld{k}\cdot\bld{v}=0$.
Thus the term involving the sound speed in the momentum equation
vanishes, and the properties of the Alfv\'en mode are independent
of the equation of energy conservation. We obtain the dispersion relation
\begin{equation}
\omega= v_A \vert k_z \vert \ ,
\end{equation}
by forming the cross product of equation \refnew{eq:eomexpand} with $\bld{k}$.
The eigenfunction satisfies
\begin{equation}
n'=\bld{k}\cdot\bld{v}=\bld{\hat{z}}\cdot\bld{v}=0 \ \ , \ \ \
\bld{b}=-{\rm sign}(k_z)\bld{v} \ .
\end{equation}
Note that both $\bld{v}$ and $\bld{b}$ are perpendicular to 
$\bld{\hat{z}}$: the Alfv\'en wave is polarized transverse to the
unperturbed magnetic field.

\subsection{Slow Mode}

We summarize the properties of the slow mode to lowest order in
$k_z/k\ll 1$, the limit appropriate to the MHD cascade.
To obtain the dispersion relation we assume, subject to verification,
that the perturbation in total pressure---i.e. thermal plus magnetic
pressure---vanishes to second order in $k_z/k$. Note that
the perturbation in total pressure is proportional to 
the terms in square brackets in both equation (\ref{eq:pcons}) and equation 
(\ref{eq:eomexpand}).
From the vanishing of this pressure term in equation (\ref{eq:eomexpand}),
we then have
\begin{equation}
\bld{k}\cdot\bld{v}= {v_A^2\over
{\tilde{c}^2+v_A^2}}
\Big[1+O\Big({k_z\over k}\Big)^2\Big]k_zv_z\ .
\label{eq:kdotv}
\end{equation}
The z-component of equation (\ref{eq:eomexpand}),
with ${\bld k}\cdot {\bld v}$
given by equation (\ref{eq:kdotv}),
then yields the dispersion relation:
\begin{equation}
\omega={\tilde{c}\over\sqrt{\tilde{c}^2+v_A^2}}
\Big[1+O\Big({k_z\over k}\Big)^2\Big]v_A \vert k_z \vert\ .
\end{equation}
Next, we solve for the eigenfunction to lowest order in $k_z/k$:
\begin{equation}
{n'\over n}\simeq {\rm sign}(k_z) {{v_Av_z}\over{\tilde{c}
\sqrt{\tilde{c}^2+v_A^2}}}
\ , \
v_x\simeq-{\tilde{c}^2\over {\tilde{c}^2+v_A^2}}
{k_z\over k} v_z \ , \ v_y=0 \ , \
\bld{b}\simeq -{\rm sign}(k_z)
{\tilde{c}\bld{v}\over\sqrt{\tilde{c}^2+v_A^2}} \ ,
\end{equation}
where the x-axis has been chosen to lie in the plane containing
$\bld{k}$ and $\bld{\hat{z}}$.
We now see that our assumption of negligible perturbed
pressure is self-consistent to lowest order in $k_z/k$.
While we have not used the x-component of equation
(\ref{eq:eomexpand}), both the left-hand side and the right-hand
side of this equation are of the same order in $k_z/k$.

Note that both $\bld{v}$ and $\bld{b}$ are nearly parallel to 
$\bld{\hat{z}}$: the slow wave polarization is nearly aligned with
the unperturbed magnetic field. Moreover, in the limit that
$\tilde{c}\gg v_A$,  the slow mode is nearly incompressible.

\subsection{Entropy Mode}

For adiabatic fluid motions, the linearized equation of energy
conservation reads
\begin{equation}
\omega s'=0 \ ,
\end{equation}
where $s$ is the entropy per particle.
Thus, there exists a zero frequency, $\omega=0$, mode with $p'= \bld{v} 
=\bld{b}=0$, which has
\begin{equation}
{T'\over T}=-{n'\over n}={\gamma-1\over \gamma}s'={2\over 5}s' \ ,
\label{eq:entropymode}
\end{equation}
with $T$ denoting temperature.
Note that there is no corresponding mode in an isothermal fluid.

\acknowledgments

\begin{center}
\begin{table}
\label{tab:lengths}
\caption{\textbf{Summary of Lengthscales}}
\begin{tabular}{ll}\hline
Symbol & Description \\ \hline
$\lambda_\perp$&lengthscale transverse to the mean magnetic field; \\
&it is comparable to the ``observed'' lengthscale.\\
$\lambda_\parallel$&lengthscale parallel to the mean magnetic field \\
$\lpar$&parallel size of an eddy; it is a function of $\lambda_\perp$
(eq. [\ref{eq:spectrum2}]). \\ \hline
$\lout$&$\lambda_\perp=\lout$ at the outer scale \\
$\lmhd$&$\lambda_\perp=\lmhd$ at the MHD scale \\
$\lcool$&$\lambda_\perp=\lcool$ at the cooling scale\\
${\rm L}_{\rm H}$&$\lambda_\perp={\rm L}_{\rm H}$ at the collisionless
scale of hydrogen atoms\\
${\rm L}_{\rm He}$&$\lambda_\perp={\rm L}_{\rm He}$ at the collisionless
scale of helium atoms\\
$\lmfp$&mean free path of protons and of electrons\\
$\led$&$\lpar=\led$ at the electron diffusion scale\\
$\lequ$&$\lpar=\lequ$ at the electron-proton equilibration scale\\
$\lpd$&$\lpar=\lpd$ at the proton diffusion scale \\
$\lpdperp$&$\lambda_\perp=\lpdperp$ at the proton diffusion scale\\
${\rm L}_{\rm p,gyr}$&$\lambda_\perp={\rm L}_{\rm p,gyr}$ at the
proton gyro scale \\ \hline\hline
\end{tabular}
\end{table}
\end{center}


\end{document}